\documentclass[prb, twocolumn,floatfix,superscriptaddress]{revtex4}

\usepackage{graphicx}

\usepackage{verbatim}

\usepackage{times}

\usepackage{subfigure}

\usepackage{amsmath}

\usepackage{natbib}


\begin{document}

\title{Pressure-energy correlations in liquids. I. Results from computer 
simulations}

\author{Nicholas P. Bailey}
\email{nbailey@ruc.dk}
\affiliation{DNRF Center ``Glass and Time'', IMFUFA, Dept. of Sciences, 
Roskilde University, P.O. Box 260, DK-4000 Roskilde, Denmark}

\author{Ulf R. Pedersen}
\affiliation{DNRF Center ``Glass and Time'', IMFUFA, Dept. of Sciences, 
Roskilde University, P.O. Box 260, DK-4000 Roskilde, Denmark}

\author{Nicoletta Gnan}
\affiliation{DNRF Center ``Glass and Time'', IMFUFA, Dept. of Sciences, 
Roskilde University, P.O. Box 260, DK-4000 Roskilde, Denmark}

\author{Thomas B. Schr{\o}der}
\affiliation{DNRF Center ``Glass and Time'', IMFUFA, Dept. of Sciences, 
Roskilde University, P.O. Box 260, DK-4000 Roskilde, Denmark}

\author{ Jeppe C. Dyre }
\affiliation{DNRF Center ``Glass and Time'', IMFUFA, Dept. of Sciences, 
Roskilde University, P.O. Box 260, DK-4000 Roskilde, Denmark}

\begin{abstract}
We show that a number of model liquids at fixed volume 
exhibit strong correlations between 
equilibrium fluctuations of the configurational parts of (instantaneous) 
pressure and energy. We present detailed results for thirteen systems, showing
in which systems these correlations are significant. These include 
Lennard-Jones liquids
(both single- and two-component) and several other simple liquids, but not 
hydrogen-bonding liquids like methanol and water, nor the Dzugutov liquid
which has significant contributions to pressure at the second nearest neighbor 
distance. The pressure-energy correlations, which for the Lennard-Jones case 
are shown to also be present in the crystal and glass phases, reflect 
an effective inverse power-law potential dominating fluctuations, 
even at zero and slightly negative pressure.
An exception
to the inverse-power law explanation is a liquid with hard-sphere repulsion and
 a square-well attractive part, where a strong correlation is observed, but
 only after time-averaging. The companion paper [arXiv:0807.0551]
 gives a thorough analysis of the correlations, with a focus on the 
Lennard-Jones liquid,
and a discussion of some experimental and theoretical consequences.
\end{abstract}

\newcommand{\angleb}[1]{\langle #1 \rangle}
\newcommand{\nod}{\noindent}
\newcommand{\half}{\frac{1}{2}}
\newcommand{\bfa}[1]{\mathbf{#1}} 

\date{\today}

\maketitle

\section{Introduction} 

\begin{figure} 
\includegraphics[width=8.5cm]{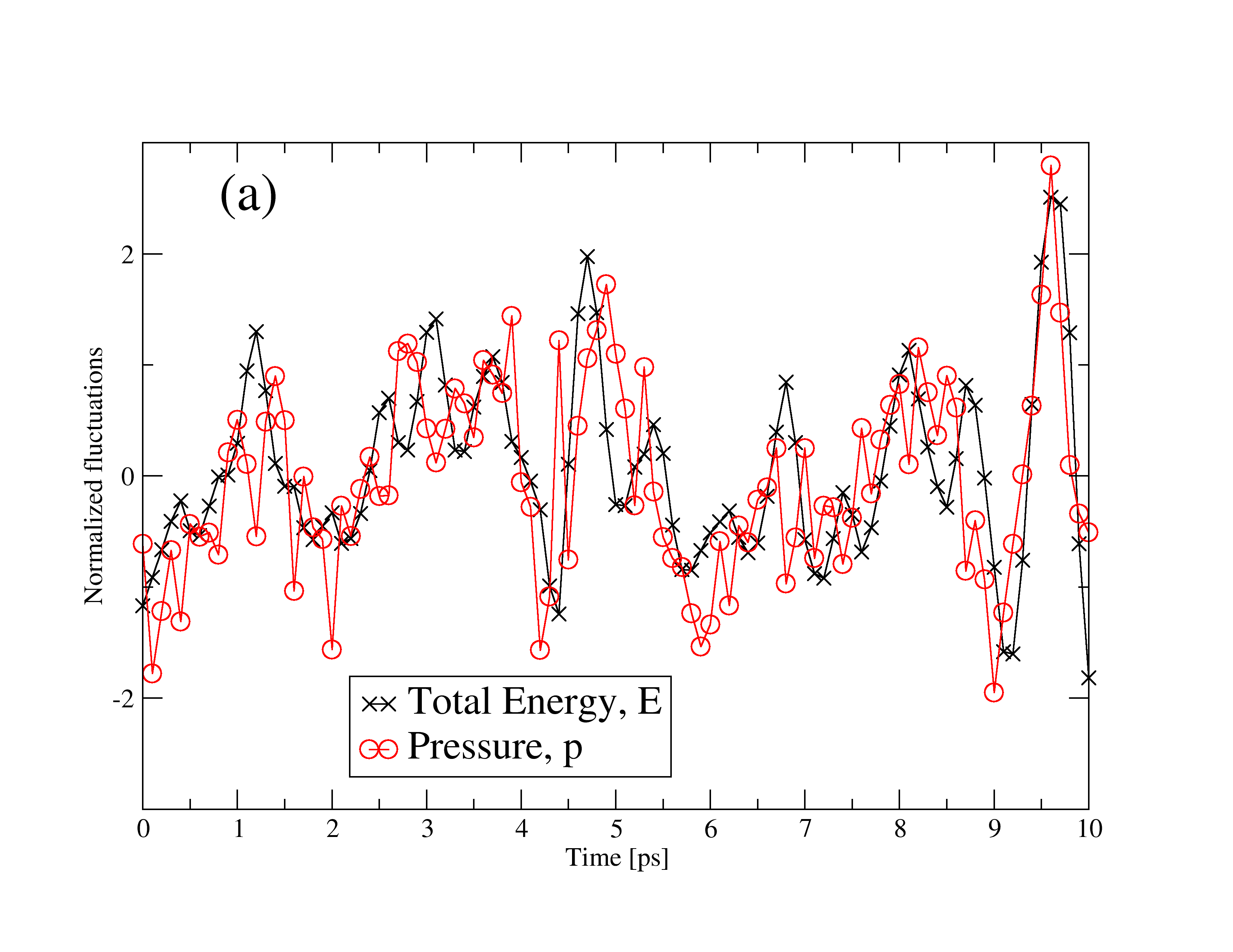} 
\includegraphics[width=8.5cm]{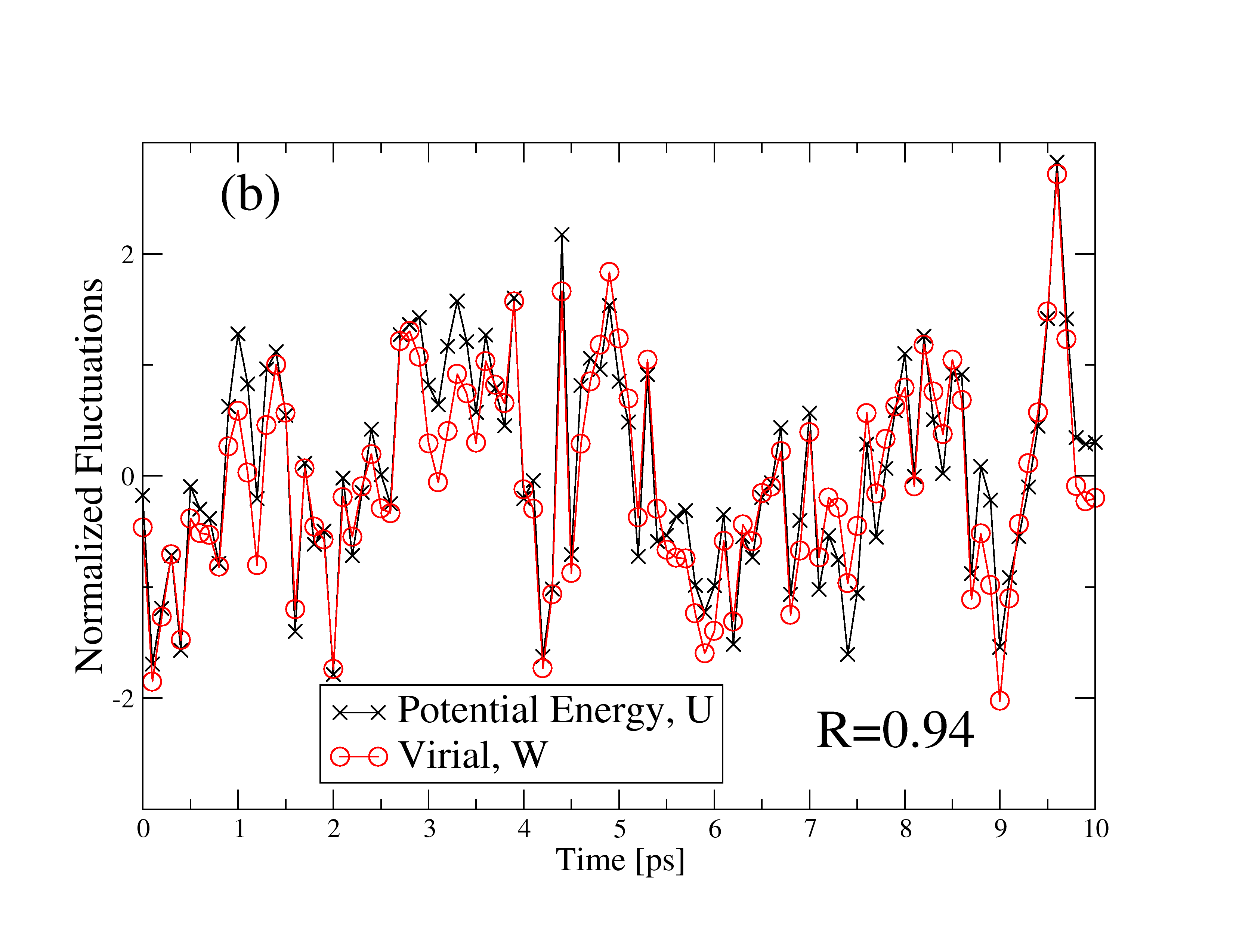} 
\caption{(Color online) Equilibrium fluctuations of (a) pressure $p$ and 
energy $E$ and (b) virial $W$ and potential energy $U$,
in a single-component Lennard-Jones system simulated in the NVT ensemble at
$\rho=34.6$ mol/l and $T=80$K (Argon units). The time-averaged pressure was 
close to zero (1.5 MPa). The correlation coefficient $R$ between $W$ and $U$ is
0.94, whereas the correlation coefficient is only 0.70 between $p$ and $E$. 
Correlation coefficients were calculated over the total simulation 
time (10 ns).}
\label{EP_timeFluc}
\end{figure}

Physicists are familiar with the idea of thermal fluctuations in equilibrium. 
They
also know how to extract useful information from them, using linear response
theory.\cite{Landau/Lifshitz:1980,Hansen/McDonald:1986,Reichl:1998,
Allen/Tildesley:1987} These 
methods started with Einstein's observation that the specific heat in the
canonical ensemble is determined by the magnitude of energy fluctuations. In 
any thermodynamic system some variables are fixed, and some fluctuate.
The magnitude of the variances of the latter, 
and of the their mutual covariance, determine the thermodynamic ``response'' 
parameters.\cite{Landau/Lifshitz:1980}  For 
example, in the canonical (NVT) ensemble, pressure $p$ and energy $E$
fluctuate; the magnitude of pressure fluctuations is related to the isothermal
bulk modulus
 $K_T\equiv - V\left(\frac{\partial p}{\partial V}\right)_T$, that of
 the energy fluctuations to the specific heat at constant volume 
$c_V\equiv T\left(\frac{\partial S}{\partial T}\right)_V$, 
while the covariance $\langle \Delta p\Delta E \rangle$ is 
related\cite{Allen/Tildesley:1987} to the thermal pressure coefficient
 $\beta_V \equiv \left(\frac{\partial p}{\partial T}\right)_V$. If the latter is
non-zero, it implies a degree of correlation between pressure and 
energy fluctuations. There is no obvious reason to suspect any particularly
 strong correlation, and to the best of our knowledge
none has ever been reported. But in the 
course of investigating the physics of highly viscous liquids by computer 
simulation, we noted strong correlations  
between pressure and energy equilibrium fluctuations in several model liquids,
also in the high temperature, low-viscosity state.
These included the most studied of all computer liquids, the Lennard-Jones
system. Surprisingly, these strong correlations survive crystallization, and 
they are also present in the glass phase. ``Strong'' here and henceforth means
a correlation coefficient of 
order 0.9 or larger. In this paper we examine several model
liquids and detail which systems exhibit strong correlations and
which do not. In the companion paper\cite{Bailey/others:2008c}
(referred to as Paper II)
we present a detailed analysis of the correlations for the single-component 
Lennard-Jones system, and discuss some consequences.

\begin{figure} 
\includegraphics[width=8.5 cm]{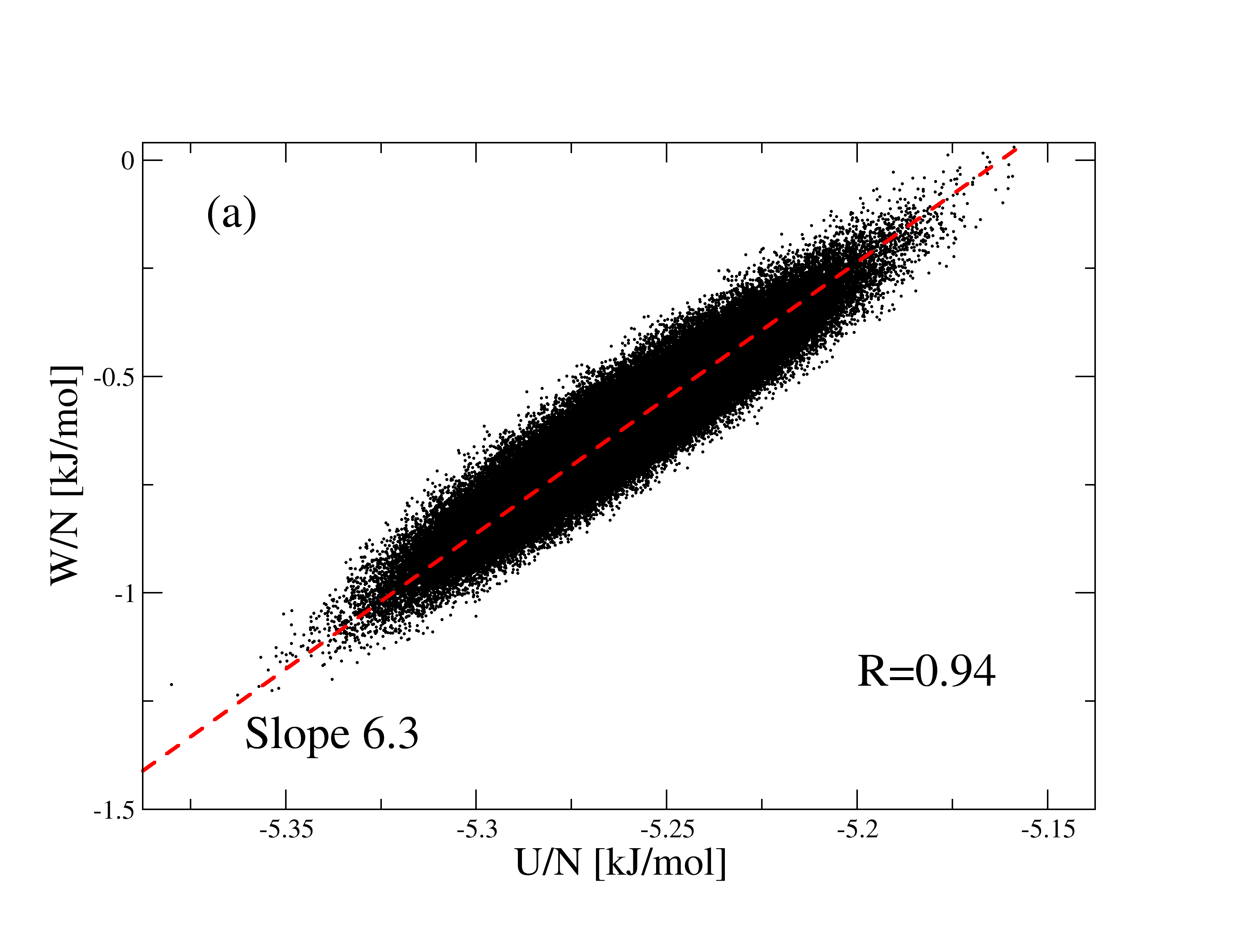} 
\includegraphics[width=8.5 cm]{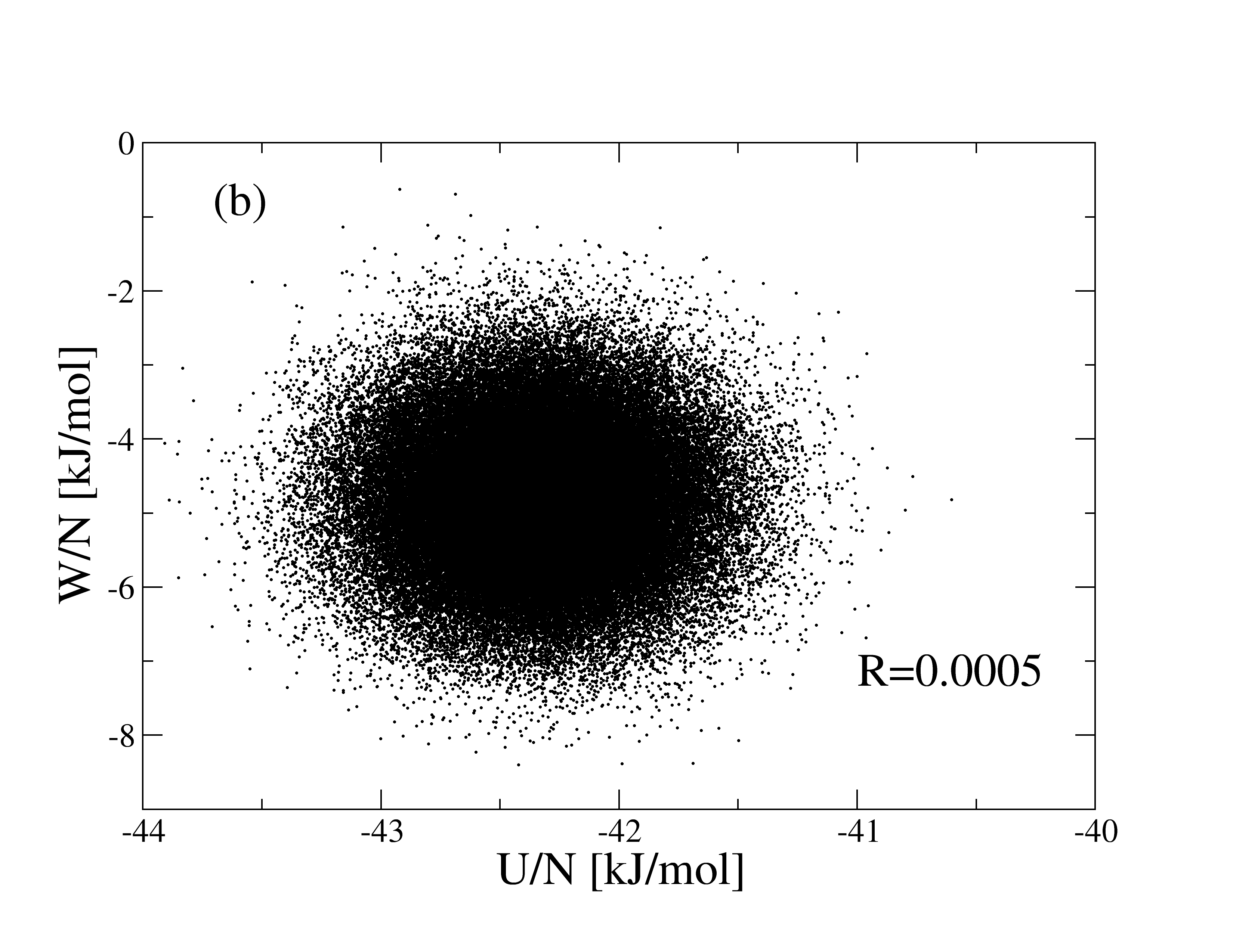} 
\caption{\label{WUscatterArgon0P80} (Color online) (a) Scatter-plot of
 instantaneous  virial $W$ and potential energy $U$ from the
 simulation of Fig.~\ref{EP_timeFluc}. The dashed line is a guide to the eye, 
with a slope determined by the ratio of standard deviations of $W$ and $U$ 
(Eq.~(\ref{slopeDefinition})). 
(b) Example of a system with almost no correlation between $W$ and $U$: 
TIP5P water at T=12.5$^\circ$C, density 1007.58 kg/m$^3$ (NVT). This system has 
Coulomb, in addition to Lennard-Jones, interactions.} 
\end{figure}

Specifically, the fluctuations which are in many cases strongly correlated are 
those of the configurational parts of pressure and energy (the parts in 
addition to the ideal gas terms). The (instantaneous)
pressure $p$ and energy $E$ have contributions both from particle momenta and
positions:

\begin{align}
 p &= Nk_BT(\bfa{p}_1,\ldots,\bfa{p}_N)/V+W(\bfa{r}_1,\ldots,\bfa{r}_N)/V \nonumber\\ 
E &= K(\bfa{p}_1,\ldots,\bfa{p}_N) + U(\bfa{r}_1,\ldots,\bfa{r}_N),
\end{align}

\nod where $K$ and $U$ and the kinetic and potential energies, respectively.
Here $T(\bfa{p}_1,\ldots,\bfa{p}_N)$ is the 
``kinetic temperature'',\cite{Allen/Tildesley:1987}
proportional to the kinetic energy per particle. The configurational 
contribution to pressure is the virial $W$, which is
 defined\cite{Allen/Tildesley:1987} by


\begin{equation}\label{generalWformula}
W = -\frac{1}{3} \sum_i \bfa{r}_i \cdot \bfa{\nabla}_{\bfa{r}_i} U
\end{equation}


\nod where $\bfa{r}_i$ is the position of the $i$th particle.
Note that $W$ has dimension energy. For a pair interaction we have

\begin{equation}
U_{\textrm{pair}} = \sum_{i<j} v(r_{ij})
\end{equation}

\nod where $r_{ij}$ is the distance between particles $i$ and $j$ and
$v(r)$ is the pair potential. The expression for the virial 
(Eq.~(\ref{generalWformula})) becomes\cite{Allen/Tildesley:1987}

\begin{equation}
W_{\textrm{pair}} = -\frac{1}{3}\sum_{i<j} r_{ij} v'(r_{ij}) = -\frac{1}{3} \sum_{i<j} w(r_{ij})
\end{equation}

\nod where for convenience we define 

\begin{equation}\label{pairVirialDef}
w(r)\equiv r v'(r).
\end{equation}

Fig.~\ref{EP_timeFluc} (a) shows normalized instantaneous values
 of $p$ and $E$, shifted and scaled to have zero mean and unit
 variance, as a function of time for the standard single-component 
Lennard-Jones (SCLJ)
liquid, while Fig.~\ref{EP_timeFluc} (b) shows the
corresponding fluctuations of $W$ and $U$. We quantify the degree of 
correlation by the standard correlation coefficient $R$, defined by

\begin{equation}\label{correlationCoeff}
R=\frac {\langle\Delta W \Delta U\rangle}{\sqrt{\langle(\Delta
    W)^2\rangle}\sqrt{\langle(\Delta U)^2\rangle}}.
\end{equation}

\nod Here angle brackets $\langle\rangle$ denote thermal averages while
$\Delta$ denotes deviation from the average value of the given quantity.
The correlation coefficient is ensemble-dependent, but our main focus---the
 $R \rightarrow1$ limit--is not. Most of the simulations reported below were 
carried out in the NVT ensemble. 
Another important characteristic quantity is the ``slope'' $\gamma$,
which we define as the ratio of standard deviations:

\begin{equation}\label{slopeDefinition}
\gamma \equiv \frac{\sqrt{(\Delta W)^2}} {\sqrt{(\Delta U)^2}}.
\end{equation}

\nod Considering the ``total'' quantities, $p$ and $E$, 
(Fig.~\ref{EP_timeFluc}(a)) there is some correlation; the 
correlation coefficient $0.70$. For the configurational parts, $W$ and $U$, 
on the other hand (Fig.~\ref{EP_timeFluc}(b)), 
the degree of correlation is much higher, $R=0.94$ in this
case. Another way to exhibit the correlation is a scatter-plot of $W$
against $U$, as shown in Fig.~\ref{WUscatterArgon0P80}(a).


Is this correlation surprising? Actually, there are some interatomic potentials
for which there is a 100\% correlation between virial and potential energy. 
If we have a pair potential of the form
 $v(r) \propto r^{-n}$, an inverse power-law, then $w(r) = -n v(r)$ and
$W_{\textrm{pair}} = (n/3) U_{\textrm{pair}}$ holds exactly. In this case the
correlation is 100\% and $\gamma=n/3$.

Conversely, suppose a system is known to be governed by a pair potential, and
 that there is 100\% correlation between $W$ and $U$.
We can write both $U$ and $W$ at any given time $t$
as integrals over the instantaneous radial distribution function
defined\cite{Allen/Tildesley:1987} as

\begin{equation}\label{timeDepRDF}
g(r,t)  \equiv \frac{2}{N \rho} \sum_{i<j} \delta(r-r_{ij}(t))/(4\pi r^2)
\end{equation}

\nod from which

\begin{equation}\label{U_from_RDF}
U(t)  = \frac{N}{2} \rho \int_0^{\infty} \! dr \, 4\pi r^2 g(r,t) v(r)
\end{equation}

\nod and 

\begin{equation}\label{W_from_RDF}
W(t)  = -\frac{N}{6} \rho \int_0^{\infty} \! dr \, 4\pi r^2 g(r,t) w(r).
\end{equation}

\begin{figure} 
\includegraphics[width=8 cm]{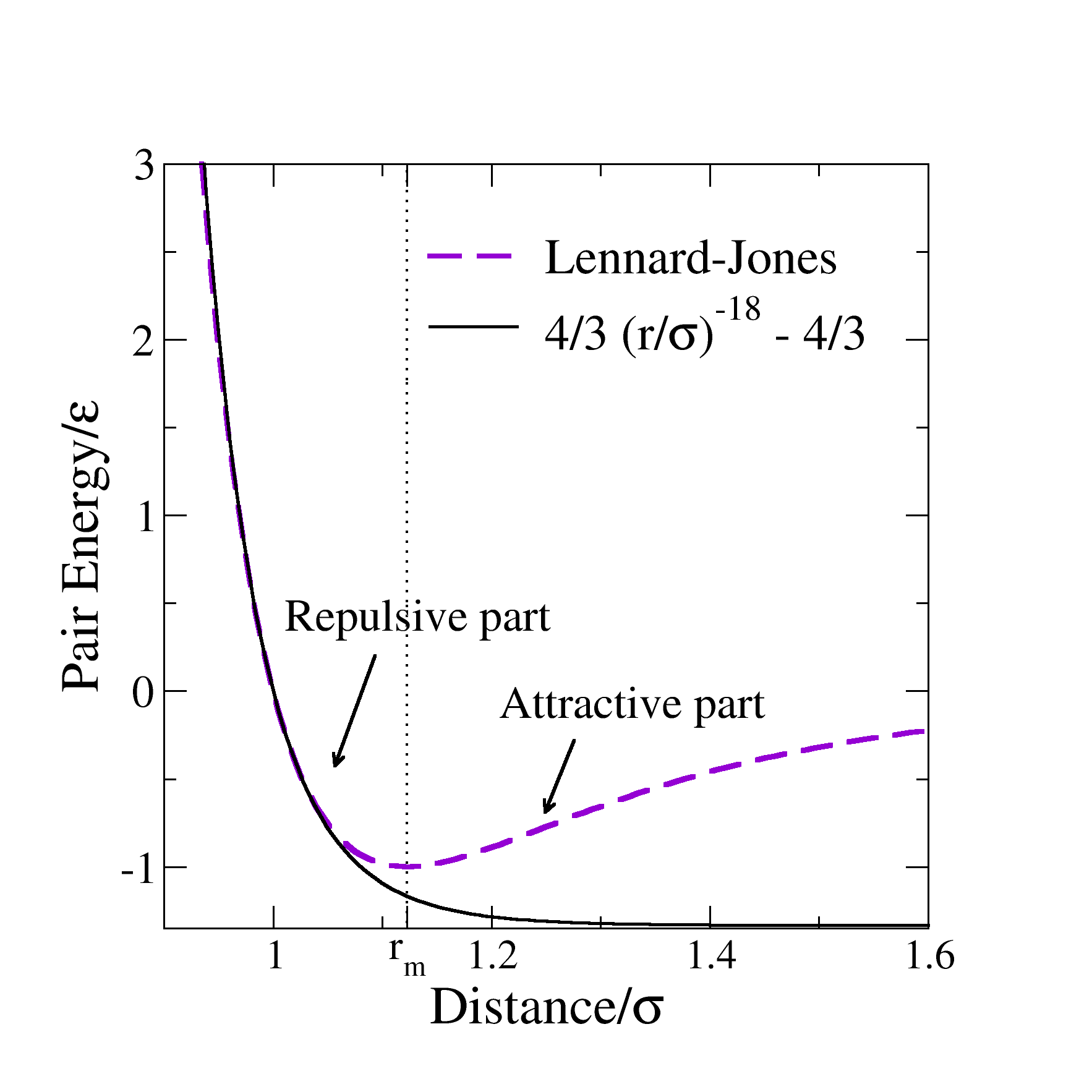} 
\caption{\label{effectivePowerLawFitSimple} (Color online) Illustration of
the ``effective inverse power-law'' chosen in this case to match the 
Lennard-Jones
 potential and its first two derivatives at the point $r=\sigma$. The vertical
 line marks the division into the repulsive and attractive parts of the 
Lennard-Jones potential.} 
\end{figure}

\nod Here the factor of $\half$ is to 
avoid double-counting, and $\rho=N/V$ is the number density. 100\% correlation 
means that $W(t) = \gamma U(t)$ holds for arbitrary $g(r,t)$ (a possible 
additive constant could be absorbed into the definition of $U$). In particular 
we
could consider $g(r,t)=\delta(r-r_0)$.\footnote{If this seems unphysical, the
argument could be given in terms of arbitrary deviations from equilibrium,
 $\Delta g(r,t)\equiv g(r,t)-\angleb{g(r)}$. See however, Paper II.} 
Substituting this into the above
expressions, the integrals go away and we find $w(r_0)=-3\gamma v(r_0)$. Since
$r_0$ was arbitrary, $v'(r) = -3\gamma v(r)/r$, which has the solution 
$v(r) \propto r^{-3\gamma}$. This connection between an inverse
 power-law potential and perfect correlations suggests that strong correlations
 can be attributed to an {\em effective inverse power-law potential}, with
 exponent given by three times the observed value of $\gamma$. This will be 
detailed in Paper II, which shows that while this explanation is basically 
correct, matters are somewhat more complicated than this. For instance,
the fixed volume condition, under which the strong correlations are actually
observed, imposes certain constraints on $g(r,t)$.

The celebrated Lennard-Jones potential is 
given\cite{Lennard-Jones:1931} by

\begin{equation}\label{LJpotential}
v_{LJ}(r) = 4\epsilon\left[ \left(\frac{\sigma}{r}\right)^{12}
  - \left(\frac{\sigma}{r}\right)^{6}  \right].
\end{equation}

\nod One might think that in the
case of the Lennard-Jones potential the fluctuations are dominated by
the repulsive $r^{-12}$ term, but this na\"{\i}ve guess leads to a slope of
four, rather than the 6.3 seen in Fig.~\ref{WUscatterArgon0P80} (a). 
Nevertheless 
the observed correlation, and the above mentioned association with inverse
power-law potentials, suggest that an effective inverse power-law
description (involving short distances), with a more careful identification
of the exponent, may apply. In fact, the presence of the second,
attractive, {\em term}, increases the effective steepness of the repulsive
{\em part}, thus increasing the slope of the correlation, or equivalently the
effective inverse
power-law exponent (Fig.~\ref{effectivePowerLawFitSimple}). 
Note the distinction between repulsive {\em term}
and repulsive {\em part} of the potential: the latter is the region 
where $v(r)$ has negative slope, thus the region  $r<r_m$ ($r_m$ being 
the distance where the pair potential has its minimum, $2^{1/6}\sigma$ for 
$v_{LJ}$). This region involves both 
the repulsive and attractive terms (see Fig.~\ref{effectivePowerLawFitSimple}, 
which also illustrates the approximation of the repulsive part by a power 
law with exponent 18).
The same division was made by Weeks, Chandler, and Andersen (WCA) in
their noted paper of 1971,\cite{Weeks/Chandler/Andersen:1971} in which they
showed that the the thermodynamic and structural properties of the 
Lennard-Jones fluid were dominated by the repulsive part at high temperatures
for all densities, and also at low temperatures for high densities.
Ben-Amotz and Stell\cite{Ben-Amotz/Stell:2003} noted that the 
repulsive core of the
 Lennard-Jones potential may be approximated by an inverse power-law with 
$n\sim$18--20. The
approximation by an inverse power-law may be directly checked by computing the 
potential and virial with an inverse power-law potential for configurations 
drawn from actual simulations using the Lennard-Jones potential. The agreement
 (apart from additive constants) is good, see Paper II. 


Consider now a system with different types of pair interactions,
for example a binary Lennard-Jones system with AA, BB, and AB
interactions, or a hydrogen-bonding system modelled via both Lennard-Jones and
Coulomb interactions. We can write arbitrary deviations of $U$ and $W$ from 
their mean values, denoted $\Delta U$ and $\Delta W$, as
a sum over types (indexed by $t$; sums over pairs of a given type
are implicitly understood):

\begin{equation}
\Delta U = \sum_t \Delta U_t;\ \Delta W = \sum_t \Delta W_t .
\end{equation}

\nod Now, supposing there is near-perfect correlation for the
individual terms with corresponding slopes $\gamma_t$, we can rewrite
$\Delta W$ as

\begin{equation}
\Delta W = \sum_t \gamma_t \Delta U_t.
\end{equation}

\nod If the $\gamma_t$ are all more or less equal to a single value $\gamma$, 
then this can be factored out and we get $\Delta W \simeq \gamma \Delta U$. 
Thus the existence of different Lennard-Jones interactions in the same
system does not destroy the correlation, since they have
$\gamma_t\sim6$. On the other hand the slope 
for Coulomb interaction, which as an inverse power-law has perfect $W,U$
correlations, is $1/3$, so we cannot expect overall strong correlation in
this case (Fig.~\ref{WUscatterArgon0P80} (b)). Indeed such reasoning also 
accounts for the reduction of 
correlation when the total pressure and energy are considered:
 $\Delta E=\Delta U+ \Delta K$, while (for a large atomic system)
$V\Delta p=\gamma\Delta U + (2/3) \Delta K$. The fact that $\gamma$ is 
(for the Lennard-Jones potential) quite different from 2/3 implies
 that the $p,E$
correlation is significantly weaker that of $W,U$ (Fig.~\ref{EP_timeFluc}).
Even in cases of unequal slopes, however, there can be circumstances under
which one kind of term, and therefore one slope, dominates the fluctuations.
In this case strong correlations will be observed. Examples include the
high-temperature limits of hydrogen-bonded liquids 
(section~\ref{ResultsAllSystems}) and the time-averaged 
(total) energy and pressure in viscous liquids (Paper II).

Some of the results detailed below were published previously in Letter 
form;\cite{Pedersen/others:2008} the aim of the present contribution is to 
make a comprehensive report covering more systems, while Paper II
contains a detailed analysis and discusses applications.
In the following section, we describe the systems simulated. In
section~\ref{Results} we present the results for all the systems
investigated, in particular the degree of correlation (correlation coefficient 
$R$) and the slope. Section~\ref{Summary} gives a summary.

\section{\label{Simulations}Simulated systems}

A range of simulation methods, thermodynamic ensembles and
computational codes were used. One reason for this was to eliminate the
possibility that the strong correlations are an artifact of using a
particular ensemble or code. In addition, no one code can simulate the
full range of systems presented. Most of the data we present are from
molecular dynamics (MD) simulations, although some are from Monte
Carlo~\cite{Landau/Binder:2005}  (MC)  and 
event-driven\cite{Zaccarelli/others:2002} (ED) simulations.
Most of the MD simulations (and of course all MC
simulations), had fixed temperature (NVT), while some
had fixed total energy (NVE). Three MD codes were used:
Gromacs 
(GRO),\cite{Berendsen/vanderSpoel/vanDrunen:1995,Lindahl/Hess/vanderSpoel:2001}
Asap (ASAP),~\cite{Asap} and DigitalMaterial (DM).~\cite{Bailey/others:2006} 
Home-made (HM) codes were used for the MC and ED simulations.

We now list the thirteen systems studied, giving each a code-name for future 
reference. The systems include monatomic systems interacting with pair 
potentials, binary atomic systems interacting with pair
potentials, molecular systems consisting  of 
Lennard-Jones particles joined rigidly together in a fixed
configuration (here the Lennard-Jones interaction
models the van der Waals forces), molecular systems which
have Coulomb as well as Lennard-Jones interactions,
 metallic systems with a many-body potential,
and a binary system interacting with a discontinuous
``square-well'' potential. Included with each system is a list
specifying which simulation method(s), which ensemble(s), and which code(s)
were used (semi-colons separate the method(s) from the ensemble(s) and the 
ensemble(s) from the code(s)). Details of the potentials are given in 
Appendix~\ref{potentialDetails}.

\begin{description}
\item[CU] Pure liquid Cu simulated using
 the many-body potential derived from effective medium theory
 (EMT); \cite{Jacobsen/Norskov/Puska:1987, Jacobsen/Stoltze/Norskov:1996}
 (MD; NVE; ASAP) 
\item[DB] Asymmetric ``dumb-bell'' molecules,\cite{Pedersen/others:2008a}
 consisting of two unlike
Lennard-Jones spheres connected by a rigid bond; (MD; NVT; GRO)
\item [DZ] The potential introduced by Dzugutov\cite{Dzugutov:1992}  as a
  candidate for a monatomic glass-forming system. Its distinguishing feature
is a peak in $v(r)$ around 1.5$\sigma$, after which it decays exponentially to
zero at a finite value of $r$; (MD; NVT, NVE; DM)

\item[EXP] A system interacting with a pair potential with exponential
repulsion and a van der Waals  attraction; (MC; NVT; HM) 
\item[KABLJ] The Kob-Andersen binary Lennard-Jones
 liquid;\cite{Kob/Andersen:1994} (MD; NVT,  NVE; GRO, DM)

\item[METH] The Gromos\cite{Scott/others:1999} 3-site model for methanol; 
(MD; NVT; GRO)

\item[MGCU] A model of the metallic alloy Mg$_{85}$Cu$_{15}$ using an EMT-based
potential; \cite{Bailey/Schiotz/Jacobsen:2004a} (MD; NVE; ASAP)
\item[OTP] A three-site model of the fragile glass-former Ortho-terphenyl 
(OTP);\cite{Lewis/Wahnstrom:1994}(MD; NVT; GRO)
\item[SCLJ] The standard single-component Lennard-Jones system with the 
interaction given in Eq.~(\ref{LJpotential}); (MD, MC; NVT, NVE; GRO, DM)
\item[SPC/E] The SPC/E model of water;\cite{Berendsen/Grigera/Straatsma:1987}
  (MD; NVT; GRO)
\item[SQW] A binary model with a pair interaction consisting of an
  infinitely hard core and an attractive square well;
  \cite{Zaccarelli/others:2002,
    Zaccarelli/Sciortino/Tartaglia:2004}(ED; NVE; HM)
\item[TIP5P] A five-site model for liquid water which reproduces 
the density anomaly.\cite{Mahoney/Jorgensen:2000} (MD; NVT; GRO)
\item[TOL] A 7-site united-atom model of toluene; (MD; NVT; GRO)

\end{description}

\nod The number of particles (atoms or molecules) was
in the range 500--2000. Particular simulation parameters ($N$, $\rho$,
$T$, duration of simulation) are given when appropriate in the results section.

\section{\label{Results}Results}

\subsection{\label{resultsSCLJ}The standard single-component Lennard-Jones system}

\begin{figure*} 
\includegraphics[width= 17.8cm]{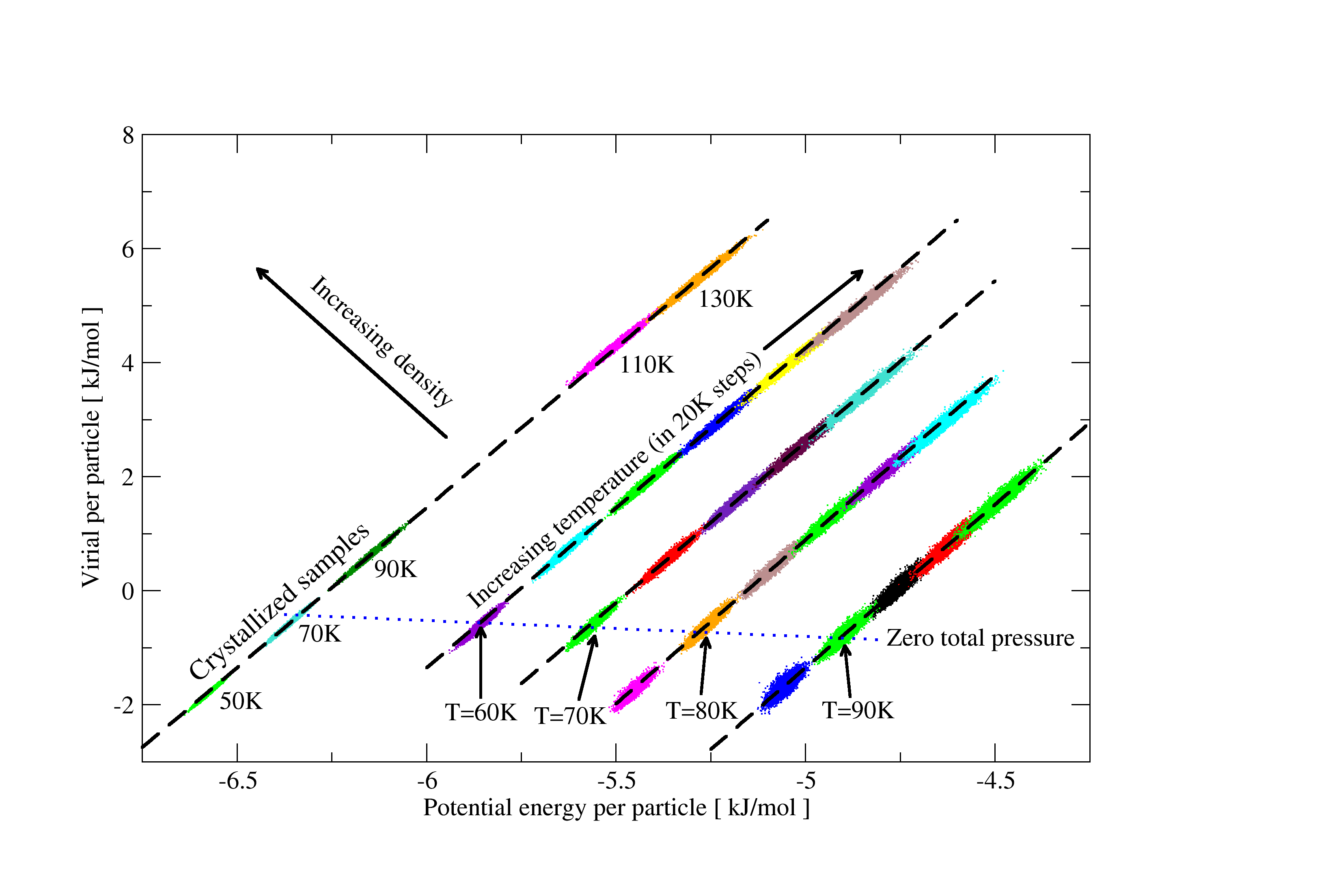} 
\caption{\label{completeArgonTP} (Color online) Scatter plots of
  the configurational parts of pressure and
  energy -- virial versus potential energy -- for several state
  points of the single-component Lennard-Jones liquid (NVT). Each oval 
  represents
  simulations at one particular temperature and density where each
  data point marks instantaneous values of virial and potential energy. The
  dashed lines mark constant density paths with the highest
  density to the upper left (densities: 39.8 mol/l, 37.4 mol/l, 36.0
  mol/l, 34.6 mol/l, 32.6 mol/l). State points on the dotted line
  have zero average pressure. The plot includes three
  crystallized samples (lower left corner), discussed at the end of
  section~\ref{resultsSCLJ} and, in more detail, in Paper II
  [reproduced from Ref.~\onlinecite{Pedersen/others:2008}].} 
\end{figure*}

\begin{table}
\caption{\label{SCLJ_correlationCoeffs} Correlation coefficients
  $R$ and effective slopes $\gamma$ for the single-component Lennard-Jones
 system (SCLJ) for the state points in Fig.~\ref{completeArgonTP}. $p$ is the 
thermally averaged pressure. The last five
states were chosen to approximately follow the isobar $p=0$.}

\begin{tabular}{|c|c|c|c|c|c|}
\hline
$\rho$ (mol/l)& $T$(K) & p(MPa) & phase & $R$ & $\gamma$ \\
\hline
42.2 &  12 &  2.6     & glass  & 0.905 &  6.02 \\
39.8 &  50 & -55.5    & crystal & 0.987 &   5.85 \\
39.8 &  70 & -0.5     & crystal & 0.989 &   5.73 \\
39.8 &  90 & 54.4     & crystal & 0.990 &  5.66 \\
39.8 & 110 & 206.2    & liquid  & 0.986 & 5.47 \\
39.8 & 150 & 309.5    & liquid  & 0.988 &  5.34 \\
37.4 &  60 & -3.7   & liquid  & 0.965 &   6.08 \\
37.4 & 100 & 102.2  & liquid  & 0.976 &   5.74 \\
37.4 & 140 & 192.7  & liquid  & 0.981 &   5.55 \\
37.4 & 160 & 234.3  & liquid  & 0.983 &   5.48 \\
36.0 &  70 & -0.7   & liquid  & 0.954 &    6.17 \\
36.0 & 110 & 90.3   & liquid  & 0.969 &    5.82 \\
36.0 & 150 & 169.5  & liquid  & 0.977 &    5.63 \\
36.0 & 190 & 241.4  & liquid  & 0.981 &    5.49 \\
36.0 & 210 & 275.2   & liquid  & 0.982 &    5.44 \\
34.6 &  60 & -42.5   & liquid  & 0.900 &    6.53 \\
34.6 & 100 & 41.7    & liquid  & 0.953 &    6.08 \\
34.6 & 140 & 114.5   & liquid  & 0.967 &   5.80 \\
34.6 & 200 & 211.0   & liquid  & 0.977 &    5.57 \\
32.6 &  70 & -35.6  & liquid  & 0.825 &    6.66 \\
32.6 &  90 & -0.8  & liquid  & 0.905 &    6.42 \\
32.6 & 110 & 31.8   & liquid  & 0.929 &    6.22 \\
32.6 & 150 & 91.7   & liquid  & 0.954 &    5.95 \\
32.6 & 210 & 172.7   & liquid  & 0.968 &    5.68 \\
\hline
37.4 & 60 & -3.7 & liquid  & 0.965 &   6.08 \\
36.0 & 70 & -0.7 & liquid  & 0.954 &   6.17 \\
34.6 & 80 & 1.5 & liquid  & 0.939 &   6.27 \\
32.6 & 90 & 0.0 & liquid  & 0.905 &  6.42 \\
42.2 & 12 & 2.6  & glass& 0.905 & 6.02 \\
\hline
\end{tabular}
\end{table}

SCLJ is the system we have most
completely investigated. $W,U$-plots are shown for a range of thermodynamic
state points in Fig.~\ref{completeArgonTP}. Here the ensemble was NVT
with N=864, and each simulation consisted of a 10 ns run taken after 10ns
of equilibration; for all SCLJ results so-called ``Argon'' units are used 
($\sigma=0.34$ nm, $\epsilon=0.997$ kJ/mol). Each elongated oval in 
Fig.~\ref{completeArgonTP} is a
collection of $W,U$ pairs for a given state point. 
Varying temperature at fixed density moves
the oval parallel to itself, following an almost straight line as indicated
by the dashed lines. Different densities
correspond to different lines, with almost the same slope.
In a system with a pure inverse power-law interaction, the
correlation would be exact, and moreover the data for all densities
would fall on the same straight line (see the discussion immediately after
Eq.~(\ref{pairVirialDef})). Our data, on the other hand,
show a distinct dependence on volume, but for a given volume, because of the
strong correlation,
the variation in $W$ is almost completely determined by that of
$U$. 

\begin{figure} 
\includegraphics[width=8.5cm]{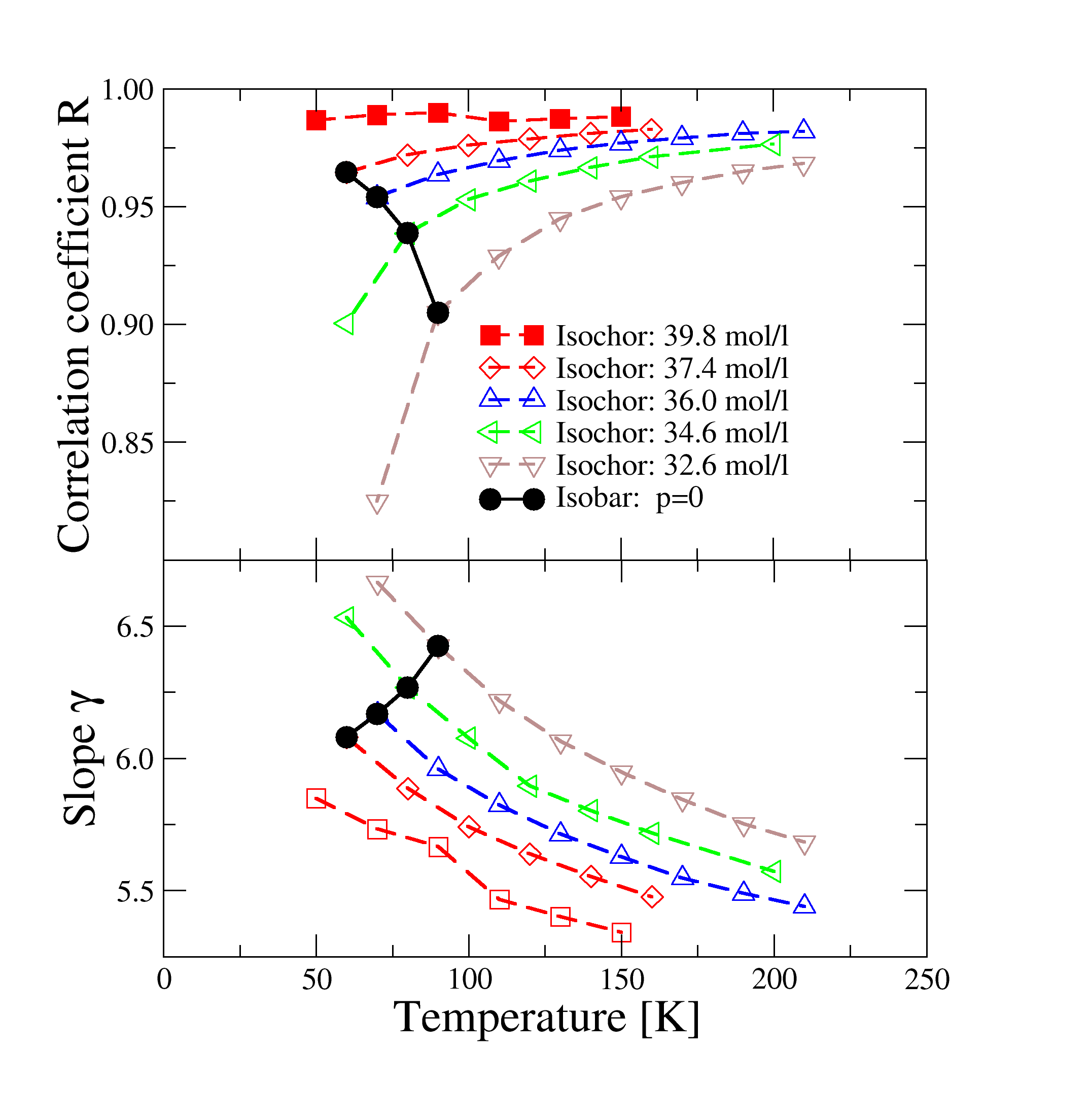} 
\caption{\label{CC_Slope_vs_T} (Color online) Upper plot, correlation 
  coefficient $R$ for the SCLJ system as a function of temperature
  for several densities (NVT). This figure makes clear the different
  effects of density and temperature on $R$. Lower plot,
  effective slope $\gamma$ as a function of $T$. Simulations at temperatures
  higher than those shown here indicate that the slope slowly approaches the 
  value four as $T$ increases. This is to be expected because as collisions
  become harder, involving shorter distances, the effective inverse
  power-law exponent approaches the 12 from the repulsive term of the 
  Lennard-Jones potential.}
\end{figure}

\nod Values of correlation coefficient $R$ for the state points of
 Fig.~\ref{completeArgonTP} are listed in
Table~\ref{SCLJ_correlationCoeffs}, along with the slope $\gamma$. 
In Fig.~\ref{CC_Slope_vs_T} we show the temperature
dependence of both $R$ and $\gamma$ for
different densities. Lines have been drawn to indicate isochores and
one isobar ($p=0$). Note that when we talk of an isobar here, we mean
a set of NVT ensembles with $V,T$ chosen so that the thermal average
of $p$ takes on a given value, rather than fixed-pressure
ensembles. This figure makes it clear that for fixed density,
$R$ increases as $T$ increases, while it also increases with density for
fixed temperature; the slope slowly decreases  in 
these circumstances. In fact it eventually reaches four, the value expected 
for a pure $r^{-12}$ interaction (e.g., at $\rho=34.6$ mol/l, $T=1000$ K,
 $\gamma=4.61$, see Ref.~\onlinecite{Pedersen/others:2008}). 
This is consistent with the idea that
 the repulsive part, characterized by an effective inverse power-law,
 dominates the fluctuations: increasing either temperature or density
increases the frequency of short-distance encounters while reducing the typical
distances of such encounters. On the other hand, along an isobar, these two
effects work against each other, since as $T$ increases, the density 
decreases. The density effect ``wins'' in this case, which
is equivalent to a statement about 
the temperature and volume derivative of $R$: Our simulations imply that

\begin{equation}
\left(\frac{\partial R}{\partial T}\right)_p  = 
\left(\frac{\partial R}{\partial T}\right)_V + 
\left(\frac{\partial R}{\partial V}\right)_T 
\left(\frac{\partial V}{\partial T}\right)_p < 0
\end{equation}

\nod which is equivalent to

\begin{equation}
\left(\frac{\partial R}{\partial T}\right)_V
 < -\left(\frac{\partial R}{\partial V}\right)_T V \alpha_p = \rho \left(\frac{\partial R}{\partial \rho}\right)_T \alpha_p
\end{equation}

\nod where $\alpha_p \equiv (\partial V/\partial T)_p/V$ is the thermal 
expansivity at constant pressure and $\rho$ is the particle density. This 
can be recast in terms of logarithmic derivatives (valid whenever 
$(\partial R/\partial \rho)_T>0$) as follows

\begin{equation}\label{rhoBeatsTempequiv}
\frac{ \left( \frac{\partial R}{\partial \ln(T)} \right)_V }
     { \left( \frac{\partial R}{\partial \ln(\rho)}\right)_T }
 < T\alpha_p.
\end{equation}

Thus what we observe in the simulations, namely that 
the correlation becomes stronger as temperature is reduced 
at fixed pressure, is a priori more to be
expected when the thermal expansivity is large (since then the right hand side 
of Eq.~(\ref{rhoBeatsTempequiv}) is large). This has 
particular relevance in the context of supercooled liquids, which we discuss in
 paper II, because these are usually studied by lowering
temperature at fixed pressure. 
On the other hand if the expansivity becomes small,
as for example, when a liquid passes through the glass transition, the
inequality (\ref{rhoBeatsTempequiv}) is a priori less likely to be satisfied.
We have in fact observed this in a simulation of OTP: upon
cooling through the (computer) glass transition, 
 the correlation became weaker with further lowering of 
temperature at constant pressure.

\begin{figure} 
\includegraphics[width=8.5cm]{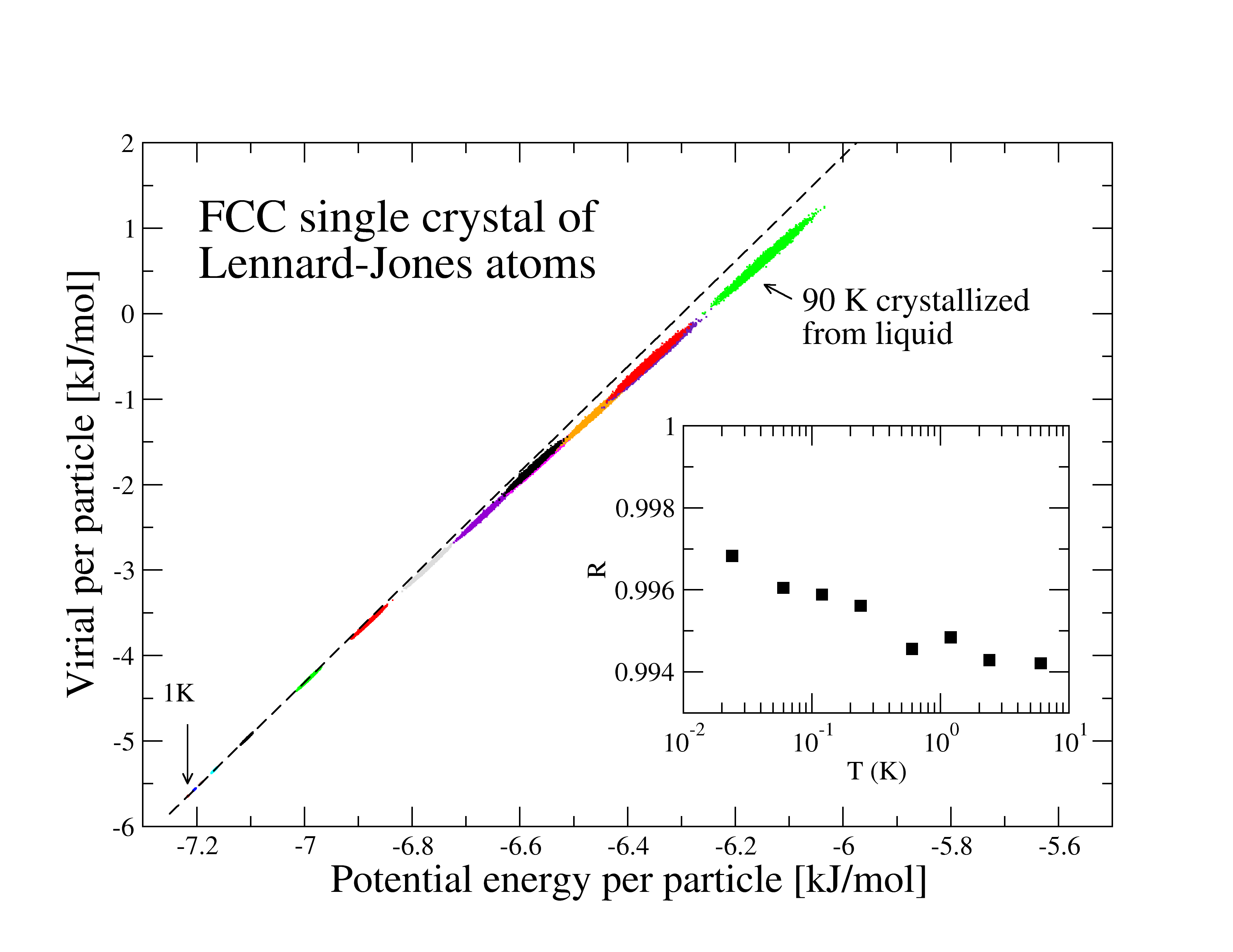} 
\caption{\label{fccArgon} (Color online) Scatter-plot of the
$W,U$ correlations for a perfect face-centered cubic 
(FCC) crystal of Lennard-Jones atoms at temperatures 1K, 2K, 3K, 5K, 10K, 20K, 
30K, 40K, 50K, 60K, 70K and 80K, as well as for defective crystals (i.e., 
crystallized from the liquid) at temperatures 50K, 70K and 90K (NVT).
The dashed 
line gives the best fit to the (barely visible) lowest-temperature data ($T=1$
K).  The inset shows the temperature dependence of $R$ at very low 
temperatures. The crystalline case is examined in detail in Paper II, where we 
find that $R$ does not converge to unity at $T=0$, but rather
to a value very close to unity.}
\end{figure}

Remarkably, the correlation persists when the system
has crystallized, as seen in the data for the highest density---the 
occurrence of the first-order phase transition can be inferred from
the gap between the data for 90K and 110K, but the data fall on the
same line above and below the transition. One would not expect the
dynamical fluctuations of a crystal, which are usually assumed to be
well-described by a harmonic approximation, to resemble those of the
high-temperature liquid. In fact for a one-dimensional crystal
of particles interacting with a harmonic potential $v(r) = \half k
(r-r_m)^2$ it is easy to show (Paper II) that there is a negative correlation 
with slope equal to -2/3. To investigate whether the harmonic approximation 
ever becomes relevant for the correlations, we prepared a perfect FCC crystal 
of SCLJ particles at zero temperature and simulated it at increasing 
temperatures, from 0.02K to 90K in Argon units, along a
constant density path. The results are shown in
Fig.~\ref{fccArgon}. Clearly the correlation is maintained right down
to zero temperature. The harmonic approximation is therefore useless for
dealing with the pressure fluctuations even as $T\rightarrow0$, 
because the slope is far from -2/3. 
The reason for this is that the dominant contribution to the 
virial fluctuations comes from the third order term, as shown in 
Paper II.

\subsection{\label{Dzugutov}A case with little correlation: the Dzugutov system}

\begin{figure} 
\includegraphics[width=8.5cm]{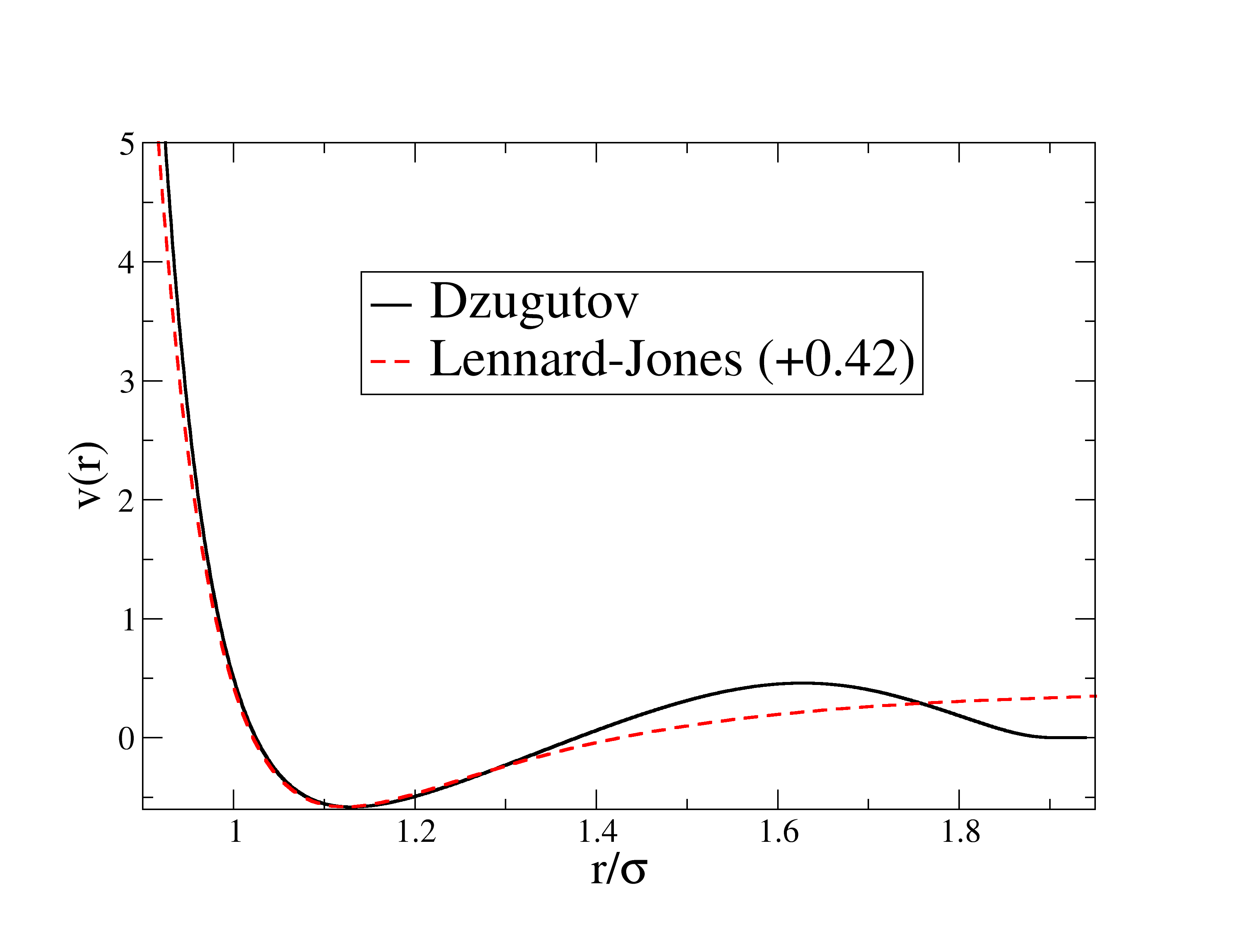} 
\caption{\label{DZpotential} (Color online) A plot of the Dzugutov pair 
potential, with the Lennard-Jones potential (shifted by a constant) shown for
comparison.}
\end{figure}

\begin{figure} 
\includegraphics[width=8.5cm]{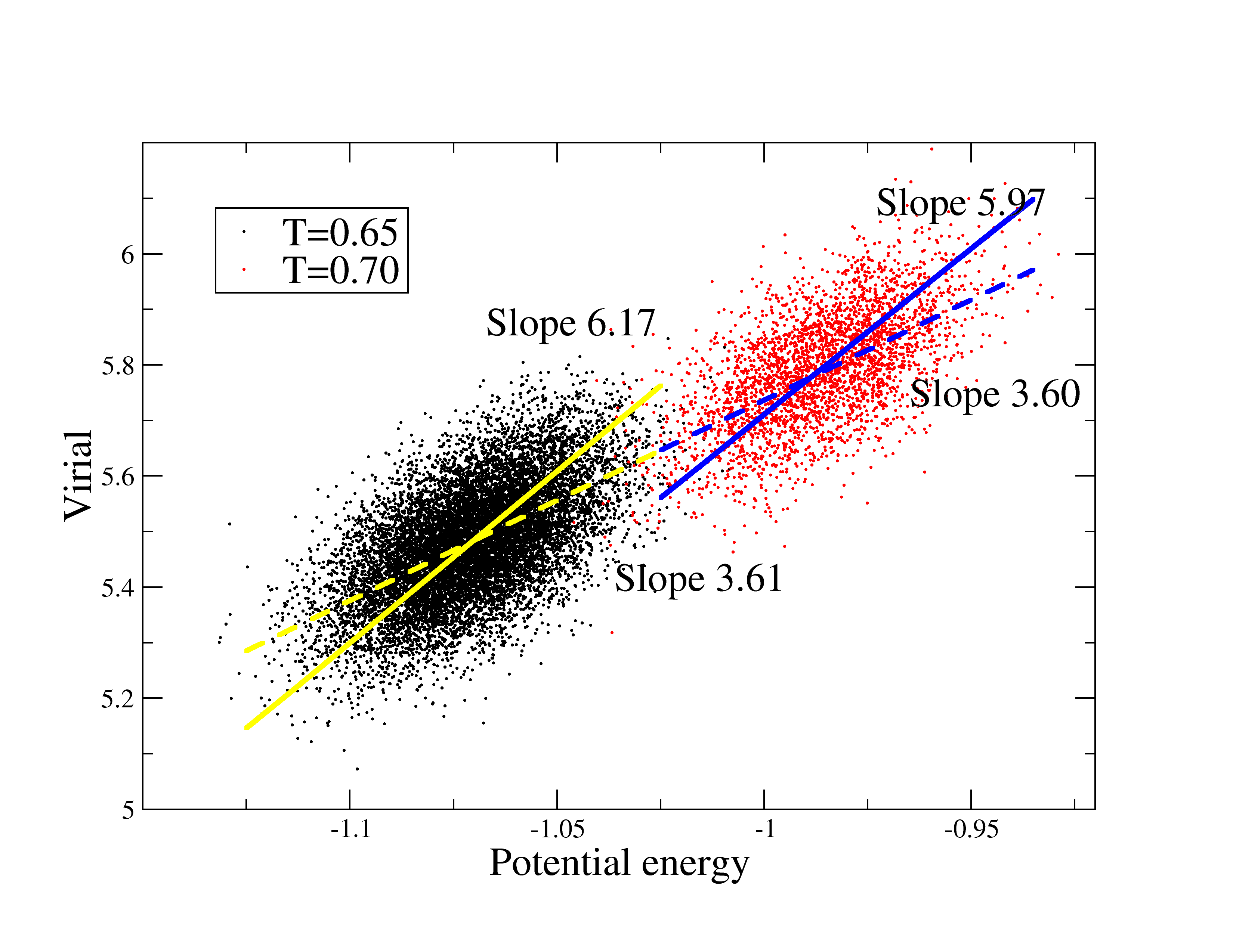} 
\caption{\label{DzugutovVPE} (Color online) Scatter-plot of 
  $W,U$ correlations for the Dzugutov
  system at density 0.88 and temperatures 0.65 and 0.70 (NVE). The
  dashed lines indicate the best-fit line using linear
  regression. These are consistent with the temperature dependence of 
  the mean  values of $\langle W\rangle$ and $\langle U\rangle$, as they
  should be (see appendix~\ref{FlucThermDerivs}), but they clearly do not 
  represent the direction of greatest
  variance. The full lines have slopes equal to the ratio of
  standard deviations of the two quantities (Eq.~(\ref{slopeDefinition})).
  The correlation coefficient is 0.585 and 0.604 for $T=0.65$ and
$T=0.70$, respectively.} 
\end{figure}

Before presenting data for all the systems studied, it is useful to
see what it means for the correlation not to hold. In this subsection we 
consider the Dzugutov system,\cite{Dzugutov:1992} whose potential contains
 a peak at the second-neighbor distance, (Fig.~\ref{DZpotential}, 
see appendix~\ref{potentialDetails} for details) whose presence might be 
expected to interfere with the effectiveness of an inverse
 power-law description. In the next subsection we show
how in a specific model of water the lack of correlation can be explicitly
seen to be the result of competing interactions. Fig.~\ref{DzugutovVPE} 
shows $W,U$ plots for the Dzugutov system
for two nearby temperatures at the same density. The ovals are much
less elongated than was the case for SCLJ, indicating a significantly weaker
correlation---the correlation coefficients here are 0.585 and
0.604, respectively. In paper II it is shown 
explicitly that the weak correlation is due to contributions arising from the
 second peak. Note that the major axes of
the ovals are not aligned with the line joining the state points, given by
the mean values of $W$ and $U$, here identifiable as
the intersection of the dashed and straight lines. On the other hand, the
lines of best fit from linear regression, indicated by the
dashed lines in each
case, {\em do} coincide with the line connecting state points. This
holds generally, a fact which follows from statistical mechanics
(appendix~\ref{FlucThermDerivs}). The interesting thing is rather that the
 major axes
point in different directions, whereas in the SCLJ case they are also 
aligned with the state-point line. The linear-regression 
slope, being equal to $\langle \Delta U \Delta W\rangle/
\langle (\Delta U)^2\rangle$, treats $W$ and $U$ in an asymmetric
manner by involving $\langle (\Delta U)^2\rangle$, but not 
$\langle (\Delta W)^2\rangle$. This is because a particular choice of 
independent and dependent variables is made. If instead we plotted $U$ 
against $W$, we would expect the slope
to be simply the inverse of the slope in the $W,U$ plot, but in fact the new
slope is $\langle \Delta U \Delta W\rangle/\langle (\Delta W)^2\rangle$. 
This equals the inverse of the original slope
only in the case of perfect correlation, where $\langle \Delta U 
\Delta W\rangle^2 = \langle (\Delta W)^2\rangle \langle (\Delta U)^2\rangle$.
For our purposes a more symmetric estimate of the slope is desired, one 
which agrees with the linear
regression slope in the limit of perfect correlation. We use simply the ratio
of standard deviations  $\sqrt{\langle(\Delta W)^2\rangle}
/\sqrt{\langle(\Delta U)^2\rangle}$ (Eq.~(\ref{slopeDefinition})). 
This slope was used to plot the dashed line in Fig.~\ref{WUscatterArgon0P80}(a)
and the full lines in Fig.~\ref{DzugutovVPE}, where it clearly represents the 
orientation of the data
better.\footnote{Choosing this measure of the slope is equivalent to 
diagonalizing the correlation matrix (the covariance matrix where the variables
 have been scaled to have unit variance) to identify the independently 
fluctuating variable. This is often done in multivariate
analysis (see, e.g., Ref.~\onlinecite{Esbensen/others:2002}), rather than 
diagonalizing the covariance matrix, when different 
variables have widely differing variances.}

\subsection{\label{TIP5Pwater}When competition between van der Waals and
  Coulomb interactions kills the correlation: TIP5P water}

As we shall see in the next section, the systems which show little correlation
include several which involve both van der Waals and Hydrogen bonding, modeled 
by Lennard-Jones and Coulomb interactions respectively. As noted already, the 
latter, being a pure inverse power-law ($n=1$), by 
itself exhibits perfect correlation with slope $\gamma=1/3$, while the 
Lennard-Jones part has near perfect correlation. But the significant difference
in slopes means that no strong correlation is seen for the full interaction. To 
check explicitly that this is the reason the correlation is destroyed we have
calculated the correlation coefficients for the Lennard-Jones and Coulomb parts
separately in a model of water. Water is chosen because the density of Hydrogen
bonds
is quite high. Simulations were done with the TIP5P model of
 water \cite{Mahoney/Jorgensen:2000} which has the feature that the density
 maximum is reasonably well reproduced. This existence of the density maximum
is in fact related to pressure and energy becoming uncorrelated, as we 
shall see.

\begin{figure} 
\includegraphics[width=8.5cm]{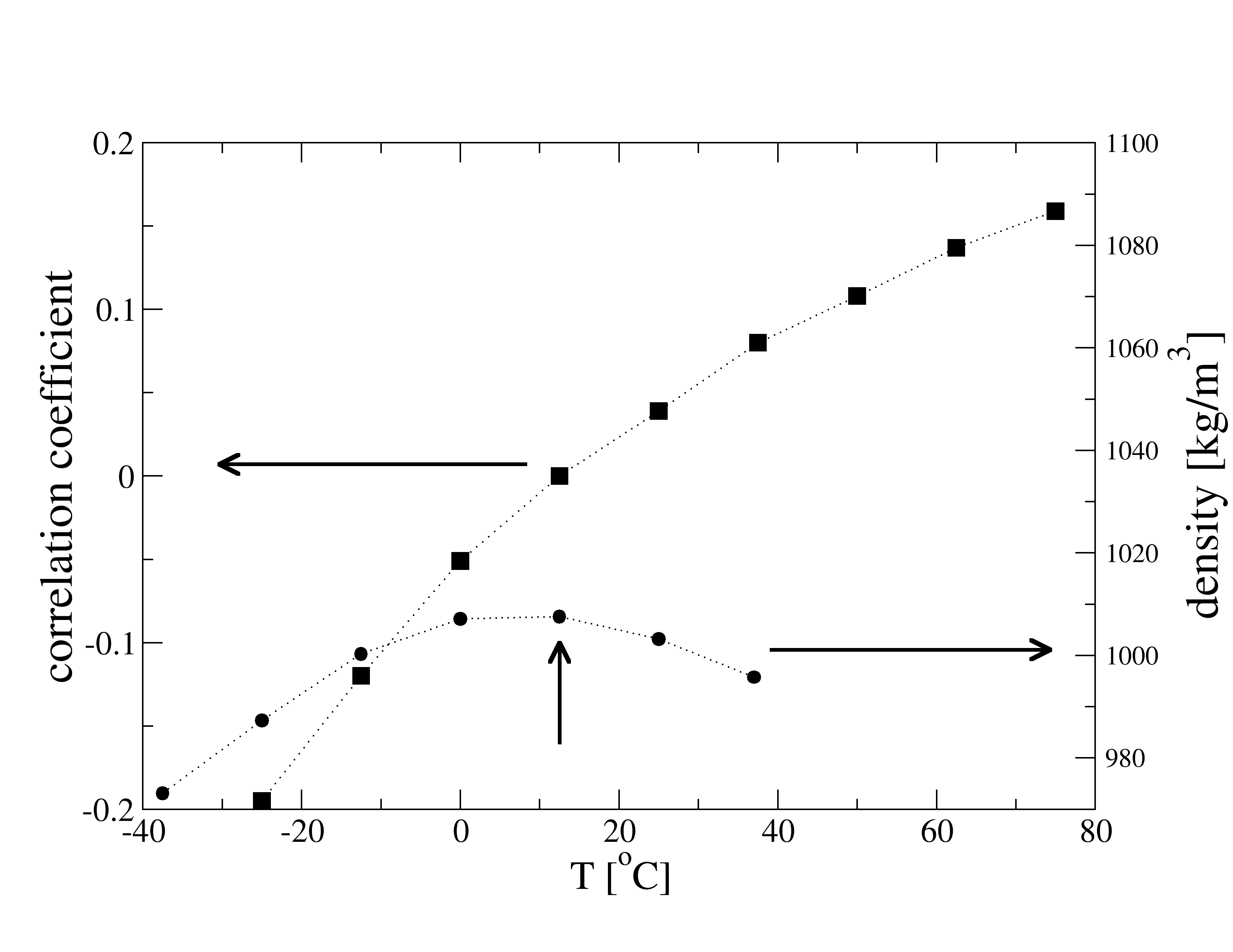} 
\caption{\label{TIP5PwaterCorrCoeff}Plot of $R$ for  TIP5P
water in NVT simulations with densities chosen to give an average pressure of
one atmosphere. Not only is the magnitude of $R$ low (less than 0.2) in 
the temperature range shown, but it changes sign around the density maximum.
The vertical arrow indicates the state point
used for Fig.~\ref{WUscatterArgon0P80} (b).
}
\end{figure}

Figure~\ref{TIP5PwaterCorrCoeff} shows the correlation coefficients and slopes
for a range of temperatures; the correlation is
almost non-existent, passing through zero around where the density attains its
maximum value. We have separately determined the correlation coefficient of
the Lennard-Jones part of the interaction; it ranges from 0.9992 at -25 
$^\circ$C to 0.9977 at 75 $^\circ$C,  even larger than
we have seen in the SCLJ system. The reason for this is that the
(attractive) Coulomb interaction forces the centers of the Lennard-Jones 
interaction closer together than they would be otherwise, thus the relevant
fluctuations are occurring higher up the repulsive part of the 
Lennard-Jones pair potential. Correspondingly the slope from this interaction 
ranges between 4.45 and 4.54, closer to the high-$T$, high density
limit of 4 than was the case for the SCLJ system. This is confirmed by
 inspection of the oxygen-oxygen radial distribution function in 
Ref.~\onlinecite{Mahoney/Jorgensen:2000} where it can be seen that the first
peak lies entirely to the left of the $v_{LJ}=0$ distance
$\sigma=0.312$ nm. Finally note that the near coincidence between the
vanishing of the correlation coefficient and the density maximum, which is
close to the experimental value of 4$^\circ$C, is not accidental: The
 correlation coefficient is proportional to the configurational part of the 
thermal pressure coefficient $\beta_V$ (Paper II). The full $\beta_V$ vanishes 
exactly
at the density maximum (4$^\circ$C), but the absence of the kinetic term means
that the correlation coefficient vanishes at a slightly higher temperature
($\sim12^\circ$C).

\subsection{\label{ResultsAllSystems}Results for all systems}

\begin{figure*} 
\includegraphics[width=15.5cm]{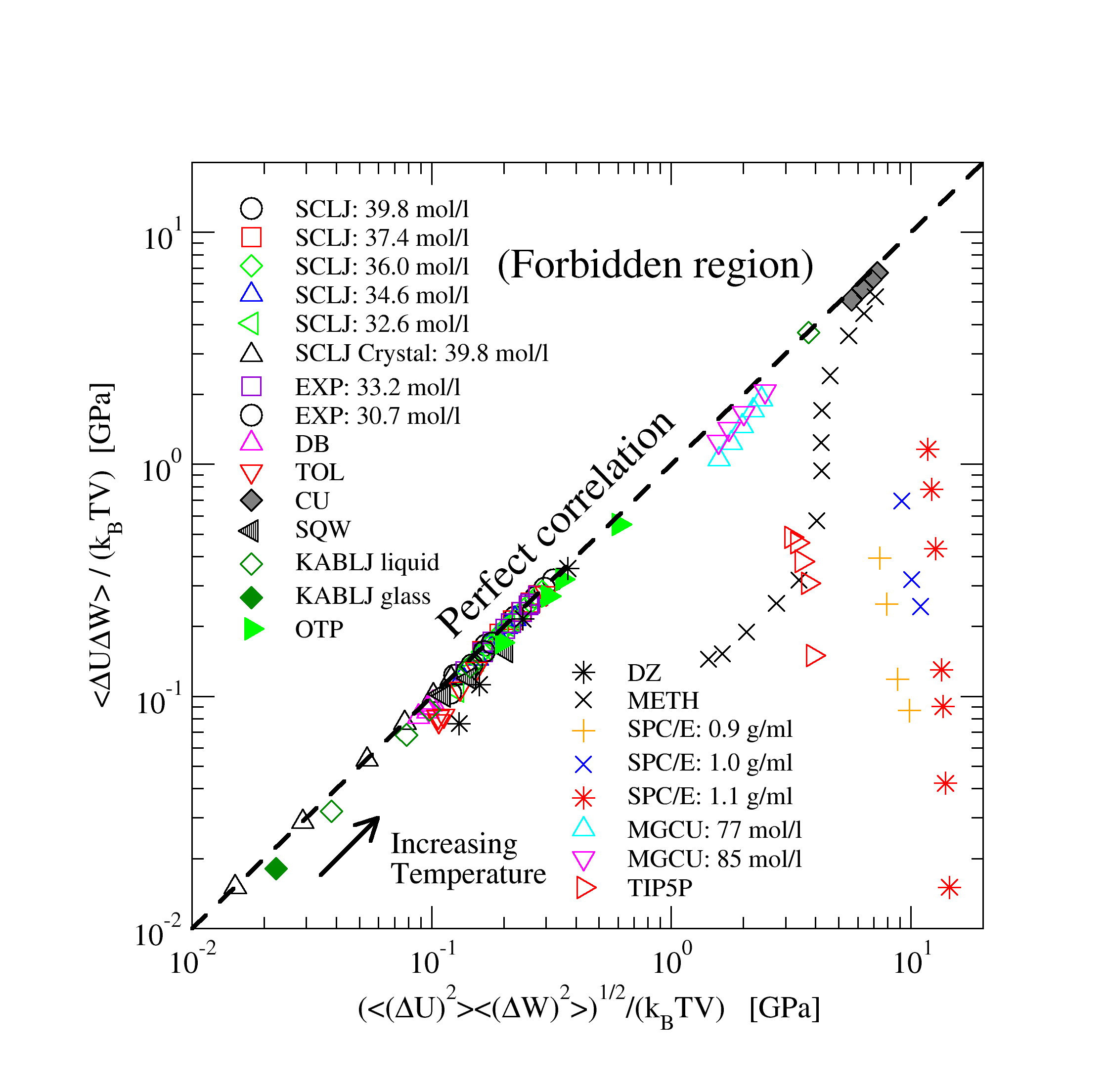} 
\caption{\label{correlationPlotAllLiquids} (Color online) $W,U$
  correlations for all simulated 
liquids; $\left< \Delta W \Delta U \right>/(k_BTV)$ plotted
  versus $\left( \left< (\Delta W)^2 \right>\left< (\Delta
      U)^2 \right> \right)^{1/2}/(k_BTV)$. Both quantities correspond to a 
pressure, which is given in units of GPa; for model systems not specifically
corresponding to real systems, such as SCLJ, KABLJ, SQW, Argon units were used 
to set the energy and length scales. If the correlation is perfect ($R=1$) the 
data fall on the diagonal marked by a dashed line. For the TIP5P model of 
water only temperatures with $R>0$ are included; 
volumes were chosen to give a pressure close to zero.} 
\end{figure*}

In Fig.~\ref{correlationPlotAllLiquids} we summarize the results for
the various systems. Here we plot the numerator of
Eq.~(\ref{correlationCoeff})
against the denominator, including factors of $1/(k_BTV)$ in both
cases to make an intensive quantity with units of
pressure. Since $R$ cannot be greater than unity, no points can appear
above the diagonal. Being exactly on the diagonal indicates perfect
correlation ($R=1$), while being significantly below it indicates poor
correlation. Different types of symbols indicate different systems, as
well as different densities for the same system, while symbols of the
same type correspond to different temperatures.



\begin{table}
\caption{\label{allSystems_correlationCoeffs}Correlation coefficients
  and effective slopes at selected state points for all investigated
  systems besides SCLJ. Argon units were used for DZ, EXP, KABLJ and SQW 
  by choosing the length parameter (of the larger particle when there were two
  types) to be 0.34 nm and the energy parameter to be 0.997 kJ/mol. 
  The phase is indicated as liquid or
  glass. SQW data involves time averaging over periods 3.0, 3.0, 8.0 
  and 9.0, respectively, for 
  the four listed state points. A minus sign has been included with the slope 
  when $R<0$; note that the $\gamma$ values only really make 
  sense as slopes when $|R|$ is close to unity. The ensemble 
  was NVT except for CU, DZ, MGCU, and SQW, where it was NVE.
}
\begin{tabular}{|c|c|c|c|c|c|}
\hline 
system & $\rho$(mol/L) & $T$(K) & phase & $R$ & $\gamma$ \\
\hline
CU     &125.8   & 1500  & liquid  & 0.907    & 4.55 \\ 
CU     &125.8   & 2340  & liquid  & 0.926    & 4.15 \\ 
DB     & 11.0   & 130   & liquid  & 0.964    & 6.77 \\
DB     & 9.7    & 300   & liquid  & 0.944    & 7.45 \\    
DZ     & 37.2   & 78    & liquid  & 0.585    & 3.61 \\ 
EXP    & 30.7   & 96    & liquid  & 0.908    & 5.98 \\
EXP    & 33.2   & 96    & liquid  & 0.949    & 5.56 \\
KABLJ  & 50.7   & 30    & glass   & 0.858    & 6.93 \\
KABLJ  & 50.7   & 70    & liquid  & 0.946    & 5.75 \\
KABLJ  & 50.7   & 240   & liquid  & 0.995    & 5.10 \\
METH   & 31.5   & 150   & liquid  & 0.318    & 22.53 \\
METH   & 31.5   & 600   & liquid  & 0.541    & 6.88 \\
METH   & 31.5   & 2000  & liquid  & 0.861    & 5.51 \\
MGCU   & 85.0   & 640   & liquid  & 0.797    & 4.74 \\ 
MGCU   & 75.6   & 465   & liquid  & 0.622    & 6.73 \\ 
OTP    & 4.65   & 300   & liquid  & 0.913    & 8.33 \\
OTP    & 4.08   & 500   & liquid  & 0.884    & 8.78 \\
OTP    & 3.95   & 500   & liquid  & 0.910    & 7.70 \\ 
SPC/E  & 50.0   & 200   & liquid  & 0.016    & 208.2 \\
SPC/E  & 55.5   & 300   & liquid  & 0.065    & 48.6 \\
SQW    & 60.8   & 48    & liquid  & -0.763   & -50.28 \\
SQW    & 60.8   & 79    & liquid  & -0.833   & -49.11 \\
SQW    & 60.8   & 120   & liquid  & -0.938   & -52.02 \\
SQW    & 59.3   & 120   & liquid  & -0.815   & -30.07 \\
TIP5P  & 55.92  & 273   & liquid  & -0.051   & -2.47 \\
TIP5P  & 55.94  & 285.5 & liquid  & 0.000    & 2.51 \\
TOL    & 10.5   & 75    & glass   & 0.877    & 7.59 \\
TOL    & 10.5   & 300   & liquid  & 0.961    & 8.27 \\
\hline 
\end{tabular}
\end{table}

All of the simple liquids, SCLJ, KABLJ, EXP, DB, TOL, 
show strong correlations, while METH, SPC/E, and TIP5P show little
correlation. Values of
$R$ and $\gamma$ at selected state points for all systems are listed in
Table~\ref{allSystems_correlationCoeffs}. What determines the degree
of correlation? The Dzugutov and TIP5P cases have already been discussed.
The poor correlation for METH and SPC/E is (presumably) because these
models, like TIP5P, involve both Lennard-Jones and Coulomb interactions.
In systems with multiple Lennard-Jones species, but without any Coulomb 
interaction, there is overall a strong
correlation because the slope is almost independent of the
parameters for a given kind of pair.

As
the temperature is increased, the data for the most poorly correlated systems,
which are all hydrogen-bonding organic molecules, slowly
approach the perfect-correlation line. This is consistent with the
idea that this correlation is observed when fluctuations of both $W$
and $U$ are dominated by close encounters of pairs of neighboring
atoms; at higher temperature there are increasingly many such
encounters, which therefore come to increasingly dominate the fluctuations. 
Also because the Lennard-Jones potential rises much more steeply
than the Coulomb potential, the latter becomes less important
as short distances dominate more. Although not obvious in the
plot, we find that increasing the density at fixed temperature generally
 increases the degree of
correlation, as found in the SCLJ case; this is also consistent with 
the increasing relevance of close encounters or collisions. 

\begin{figure}
\includegraphics[width=8.5 cm]{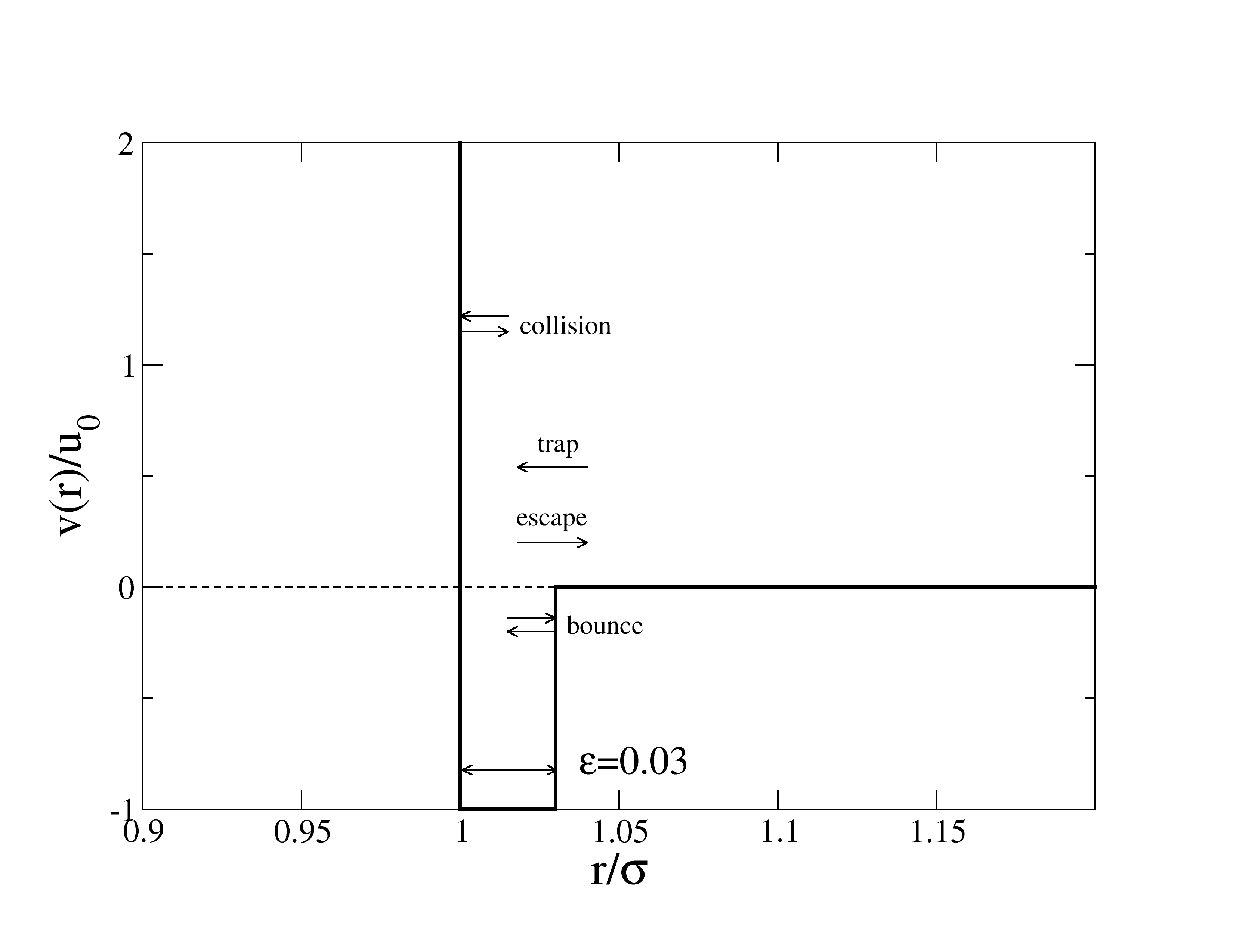}
\caption{\label{SQW_Illustration}Illustration of the square-well potential,
indicating the four microscopic processes which contribute to the virial.}
\end{figure}

\begin{figure}
\includegraphics[width=8.5 cm]{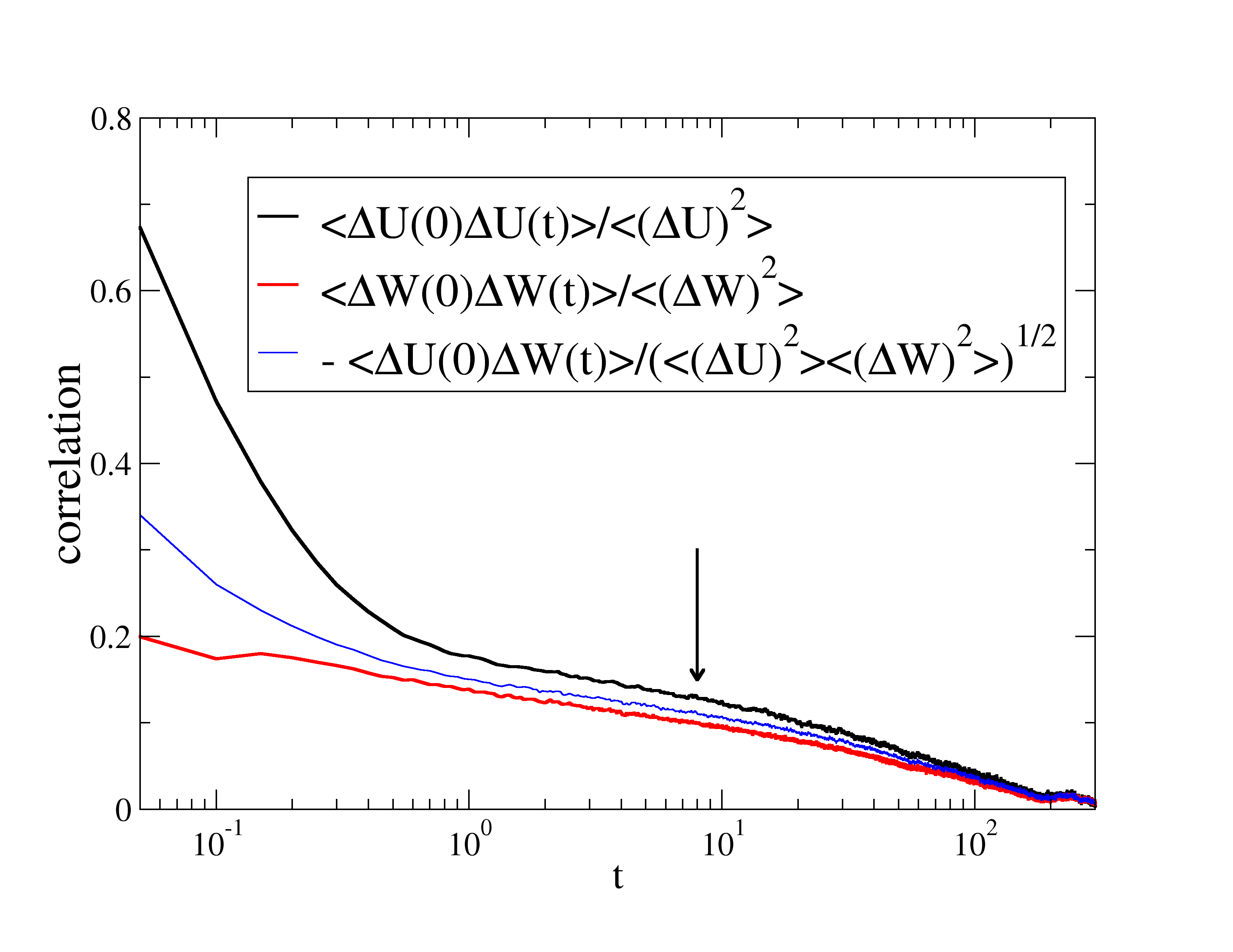}
\caption{\label{SQW_correlationFuncs} (Color online) Energy-energy, 
virial-virial, and energy-virial correlation functions for SQW at packing
 fraction $\phi=0.595$ and temperature $T=1.0$ (normalized to unity at $t=0$). 
The cross correlation has been multiplied by -1. The arrow marks the time 
$t=8$, roughly 1/10 of the relaxation time (determined from the long-time part
of the energy-virial cross-correlation function). This time was used for 
averaging.}
\end{figure}

A system quite different from the others presented
so far is the binary square-well system, SQW, with a discontinuous
potential combining hard-core repulsion and a narrow attractive well 
(Fig.~\ref{SQW_Illustration}; see 
appendix~\ref{potentialDetails} for details). Such a potential
models attractive colloidal systems,\cite{Zaccarelli/others:2002} one of whose
interesting features, predicted from simulations and theory, is the existence
of two different glass phases, termed the ``repulsive'' and ``attractive''
glasses.~\cite{Sciortino:2002} The discontinuous potential not only 
makes the simulations substantially different
from a technical point of view, but there are also conceptual
differences---in particular, the instantaneous virial is a sum of 
delta-functions in time. The dynamical algorithm employed in the
simulations is ``event-driven'', where events involve a change in the relative
 velocity of a pair of particles due to hitting the hard-core inner wall of the
potential or crossing the potential-step. The algorithm must detect the next
event, advance time by the appropriate amount, and adjust the velocities of all
particles appropriately. There are four kinds of events
 (illustrated in Fig.~\ref{SQW_Illustration}): (1) 
``collisions'', involving the inner repulsive core; (2) ``bounces'', involving
bouncing off the outer (attractive) wall of the potential; 
(3) ``trapping'', involving the separation going
below the range of the outer wall and (4) ``escapes'', involving the separation
increasing beyond the outer wall. To obtain meaningful values of the virial a 
certain amount of time-averaging must be done---we can no longer consider truly
instantaneous quantities. As shown in appendix~\ref{Virial_SQW} the 
time-averaged virial may be written as the following sum over all events which
take place in the averaging interval $ t_{\textrm{av}}$:

\begin{equation}\label{SQWvirial}
\bar{W}  = \frac{1}{3 t_{\textrm{av}}} \sum_{\textrm{events}\, e} m_{r,e}\bfa{r}_{e} \cdot \Delta \bfa{v}_{e}
\end{equation}

\nod Here $\bfa{r}_e$ and $\bfa{v}_e$ refer to the relative position and 
velocity for 
the pair of particles participating in event $e$, while $\Delta$ indicates the
change taking place in that event. Positive contributions to $\bar{W}$
come from collisions; the three other event types involve the outer wall, 
which, 
as is easy to see, always gives a negative contribution. The default
$t_{\textrm{av}}$ in the simulation was 0.05. Strong correlations emerge only at
 longer averaging times, however. An appropriate  $t_{\textrm{av}}$
may be chosen by considering the correlation functions  (auto- and cross-) for
virial and potential energy, plotted in Fig.~\ref{SQW_correlationFuncs}, where 
the ``instantaneous'' values $E(t)$ and $W(t)$ correspond to averaging over
 0.05 time units. We choose $t_{\textrm{av}} \simeq \tau_\alpha/10$, where
 $\tau_\alpha$ is the relaxation time determined from the cross-correlation
function $\langle\Delta U(0)\Delta W(t)\rangle$. 
Data for a few state points are
shown in Table~\ref{allSystems_correlationCoeffs}. Remarkably, this 
system, so different from the continuous potential systems,
 also exhibits strong $W,U$-correlations, with $R=0.94$ in the case
$T=1.0, \phi=0.595$ (something already 
hinted at in Fig.~\ref{SQW_correlationFuncs} in the fact that the curves 
coincide). There is a notable difference, however, compared to continuous
systems: The correlation is negative. 

The reason for the negative correlation is that at high density, most
of the contributions to the virial are from collisions: a particle will collide
with neighbor 1, recoil and then collide with neighbor 2 before there is a 
chance to make a bounce event involving neighbor 1. The number of collisions
that occur in a given time interval is proportional to the number of bound
pairs which is exactly anti-correlated with the energy. The effective 
slope $\gamma$ has a large (negative) value of order -50, which does not seem
to depend strongly on temperature. This example is interesting because it shows
 that strong pressure-energy 
correlations can appear in a wider range of systems that might first have been
guessed. Note, however, that the ordinary hard-sphere system cannot display 
such correlations, since potential energy does not exist as a dynamical 
variable in this system, i.e., it is identically zero. The idea of 
correlations emerging when quantities are 
averaged over a suitable time interval is one we shall meet again
in Paper II in the context of  viscous liquids.

\section{\label{Summary}Summary}

We have demonstrated a class of model liquids
whose equilibrium thermal fluctuations of virial and potential energy
are strongly correlated.  We have presented detailed
investigations of the presence or absence of such correlations in
various liquids, with extra detail presented for the standard single-component
 Lennard-Jones case. One notable aspect is how widespread these correlations
are, appearing not just in Lennard-Jones potentials or potentials
which closely resemble the Lennard-Jones one, but also in systems
involving many-body potentials (CU, MGCU) and discontinuous potentials
(SQW). We have seen how the presence of 
different types of terms in the potential, such as
Lennard-Jones and Coulomb interactions, spoil the correlations, even though
each by itself would give rise to a strongly correlating system. Most 
simulations were carried out in the NVT ensemble; $R$ is  
ensemble dependent, but the $R\rightarrow 1$ limit is not.

 Several of the hydrocarbon liquids studied here were simulated using
simplified ``united-atom'' models where groups such as methyl-groups or even 
benzene rings were represent by Lennard-Jones spheres. These could also be 
studied using more realistic ``all-atom'' models, where every atom (including
hydrogen atoms) is included.
It would be worth investigating whether the strength of the correlations is
reduced by the associated Coulomb terms in such cases.

In paper II we provide a detailed analysis for the single-component
Lennard-Jones case, including consideration of contributions beyond the 
effective inverse power-law approximation and the $T\rightarrow0$ limit of
the crystal. There we also discuss some 
consequences, including a direct experimental verification of the 
correlations for supercritical Argon and consequences of strong 
pressure-energy correlations in highly viscous liquids and biomembranes.

\begin{acknowledgments}
Useful discussions with S{\o}ren Toxv{\ae}rd are gratefully acknowledged.
Center for viscous liquid dynamics ``Glass and Time'' is sponsored by The 
Danish National Research Foundation.
\end{acknowledgments}

\appendix

\section{\label{potentialDetails}Details of interatomic potentials}

Here we give more detailed information about the interatomic potentials
 used. These details have been published elsewhere as indicated,
 except for the case of EXP and TOL.

\begin{description}
\item[CU] Pure liquid Cu simulated using
 the many-body potential derived from effective medium theory
 (EMT). \cite{Jacobsen/Norskov/Puska:1987, Jacobsen/Stoltze/Norskov:1996}
This is similar to the embedded atom method of Daw and 
Baskes,\cite{Daw/Baskes:1984} where
the energy of a given atom $i$, $E_i$ is some nonlinear function (the 
``embedding function'') of the electron density due to the neighboring atoms. 
In the EMT, it is given as the energy of an atom in an equivalent 
reference system, the ``effective medium'', plus a correction term, 
$E_i=E_{C,i}(n_i) +\half\left[\sum_{j\neq i} v_{ij}(r_{ij}) - 
\sum_{j\neq i}^{\textrm{ref}}
 v_{ij}(r_{ij})\right]$. Specifically, the reference system is chosen as an FCC
 crystal of the given kind of atom, and ``equivalent'' means that the electron
 density is used to choose the lattice constant of the crystal. The correction 
term is an ordinary pair potential involving a simple exponential, but notice 
that the corresponding sum in the reference system is
subtracted (guaranteeing that the correct reference energy is given
when the configuration is fact, the reference configuration). 
The parameters were $E_0$=-3.510 eV; $s_0$=1.413\AA ; $V_0$=2.476 eV; $\eta_2$=
3.122\AA$^{-1}$; $\kappa$=5.178; $\lambda$=3.602; $n_0$=0.0614 \AA$^{-3}$.
\item[DB] Asymmetric ``dumb-bell'' molecules,\cite{Pedersen/others:2008a}
 consisting of two unlike Lennard-Jones spheres, labelled P and M,
 connected by a rigid bond. The parameters were
$\epsilon_p$=5.726 kJ/mol, $\sigma_p=$0.4963 nm, $m_p$=77.106 u; 
$\epsilon_m$=0.66944 kJ/mol, $\sigma_m=$0.3910 nm, $m_m$=15.035 u; the bond 
length
was $d$=0.29 nm. Cross interactions, $\epsilon_{pm}$ and $\sigma_{pm}$, were set
 equal to the geometric and arithmetic means of the $p$ and $m$
parameters, respectively (Lorentz-Berthelot mixing rule\cite{Rowlinson:1969}).
\item [DZ] A monatomic liquid introduced by Dzugutov as a
  candidate for a monatomic glass-forming system.\cite{Dzugutov:1992}
 The potential is a sum of two parts, 
$v(r)=v_1(r)+v_2(r)$, with $v_1(r) = A(r^{-m}-B)\exp(c/(r-a))$ for $r<a$
and zero otherwise, and $v_2(r)=B \exp(d/(r-b))$ for $r<b$, zero 
otherwise.
The parameters are chosen to match the location and curvature of the 
Lennard-Jones potential:  $m=16$, $A=5.82$, $c=1.1$, $a=1.87$,
$B=1.28$, $d=0.27$, $b=1.94$.
\item[EXP] A system interacting with a pair potential with exponential
repulsion $U(r) = \frac{\epsilon}{8} \left[6\exp(-14(r/\sigma -
    1))-14(\sigma/r)^6\right]$. Note that the attractive term has the same
form as the Lennard-Jones potential.
\item[KABLJ] The Kob-Andersen binary Lennard-Jones
 liquid\cite{Kob/Andersen:1994}, a
mixture of two kinds of particles A and B, with A making 80\% of the total 
number. The energy and length parameters are $\epsilon_{AA}=1.0$, 
$\epsilon_{BB}=0.5$, $\epsilon_{AB}=1.5$, $\sigma_{AA}=1.0$, $\sigma_{BB}=0.88$,
 $\sigma_{AB}=0.8$. The masses are both equal to unity. We use the standard 
density $\rho=1.2\sigma_{AA}^{-3}$.

\item[METH] The Gromos 3-site model for 
methanol.\cite{vanGunsteren/others:1996}
 The sites represent the methyl (M$\equiv$CH$_3$) 
group ($m=15.035$ u), the O atom 
($m=15.999$ u), and the O-bonded H atom ($m=1.008$ u). The charges for Coulomb 
interactions are respectively 0.176 e, -0.574 e, and 0.398 e. The M and O
groups additionally interact via Lennard-Jones forces, with parameters 
$\epsilon_{MM} = 0.9444$ kJ/mol, $\epsilon_{OO} = 0.8496$ kJ/mol, 
$\epsilon_{MO} = 0.9770$ kJ/mol,
 $\sigma_{MM} = 0.3646$ nm, 
$\sigma_{OO} = 0.2955$ nm, and $\sigma_{MO} = 0.3235$ nm. Lennard-Jones 
interactions are smoothly cutoff between 0.9 nm and 1.1 nm. The M-O and O-H
distance is fixed at 0.136 nm and 0.1 nm, respectively, while the M-O-H bond
angle is fixed at 108.53$^\circ$.

\item[MGCU] A model of the metallic alloy Mg$_{85}$Cu$_{15}$, simulated
by EMT with parameters as in Ref.~\onlinecite{Bailey/Schiotz/Jacobsen:2004a}.
In this potential there are seven parameters for each element. 
However, some of the Cu parameters were 
allowed to vary from their original values in the process of optimizing the
potential for the Mg-Cu system. The parameters for Cu were 
$E_0$=-3.510 eV; $s_0$=1.413\AA ; $V_0$=1.994 eV; $\eta_2$=
3.040\AA$^{-1}$; $\kappa$=4.944; $\lambda$=3.694; $n_0$=0.0637 \AA$^{-3}$, while 
those for Mg were $E_0$=-1.487 eV; $s_0$=1.766\AA ; $V_0$=2.230 eV; $\eta_2$=
2.541\AA$^{-1}$; $\kappa$=4.435; $\lambda$=3.293; $n_0$=0.0355 \AA$^{-3}$. It
should be noted that there is an error in 
Ref.~\onlinecite{Bailey/Schiotz/Jacobsen:2004a}: The parameter $s_0$ for Cu is 
given in units of bohr instead of \AA.

\item[OTP] The Lewis-Wahnstr{\"o}m three-site model of
 orthoterphenyl\cite{Lewis/Wahnstrom:1994} consisting of three identical
Lennard-Jones spheres located at the apices A B, and C of an isosceles 
triangle. Sides AB and BC are 0.4830 nm long, while the ABC angle is 75$^\circ$.
The Lennard-Jones interaction parameters are
$\epsilon=4.989$ kJ/mol, $\sigma=0.483$ nm, while the mass of each sphere, not
specified in Ref.~\cite{Lewis/Wahnstrom:1994}, was taken as one third of the
mass of an OTP molecule, $m=76.768$ u.

\item[SCLJ] The standard single-component Lennard-Jones system with potential 
given by Eq.~(\ref{LJpotential}).

\item[SPC/E] The SPC/E model of water,\cite{Berendsen/Grigera/Straatsma:1987}
 in which each molecule consists of three rigidly bonded
point masses, with an OH distance of 0.1 nm and the HOH angle
equal to the tetrahedral angle. Charges on O and each H are
equal to -0.8476 e and +0.4238 e, respectively. O atoms interact with 
each
other via a Lennard-Jones potential with $\epsilon$=2.601 kJ/mol and
 $\sigma$=0.3166 nm.

\item[SQW] A binary model with a pair interaction consisting of an
  infinitely hard core and an attractive square well:
  \cite{Zaccarelli/others:2002,
    Zaccarelli/Sciortino/Tartaglia:2004} $v_{ij}(r) = \infty$, $r 
< \sigma_{ij}$, $v_{ij}(r)=-u_0$, 
$\sigma_{ij}<r<\sigma_{ij}(1+\epsilon)$, $v_{ij}(r)=0$, 
$r>\sigma_{ij}(1+\epsilon)$. The radius parameters are $\sigma_{AA}=1.2$,
$\sigma_{BB}=1$, $\sigma_{AB}=1.1$, while $\epsilon=0.03$ and $u_0=1$. The 
composition was equimolar, and the masses of both particles were equal to
unity.
\item[TIP5P] In this water model\cite{Mahoney/Jorgensen:2000} 
there are five sites associated with a single
water molecule. One for the O atom, one for each H, and two to locate the
centers of negative charge corresponding to the electron lone-pairs on the O. 
The OH bond length, and HOH bond angle are
fixed at the gas-phase experimental values, $r_{OH}=0.09572$ nm and
 $\theta_{HOH}=104.52^\circ$. The negative charge sites are located symmetrically
along the lone-pair directions at distance $r_{OL}=0.07$ nm and 
with an intervening angle 
$\theta_{LOL}=109.47^\circ$. A charge of +0.241 $e$ is located on each Hydrogen 
site, while charges of equal magnitude and opposite sign are placed on the
lone-pair sites. O atoms on different molecules interact via the Lennard-Jones
potential with $\sigma_O=0.312$ nm and $\epsilon_O=0.669$ kJ/mol.

\item[TOL] A 7-site united-atom
model of toluene, consisting of six ``ring'' C atoms and 
a methyl group (H atoms are not explicitly represented). In order to 
handle the constraints more easily, only three mass points were used; one at
the ring C attached to the methyl group ($m=40.065$ u), and one at each of
the two ``meta'' C atoms ($m=26.038$) (note that with this mass distribution, 
the moment of inertia is not reproduced correctly). Parameters were derived 
from the information in Ref.~\onlinecite{Jorgensen/Madura/Swenson:1984}: 
$\epsilon_{ring}$=0.4602 kJ/mol, $\epsilon_{methyl}$=0.6694 kJ/mol, 
$\sigma_{ring}$=0.375 nm, $\sigma_{methyl}$=0.391 nm. The Lorentz-Berthelot 
rule was used for cross-interactions.\cite{Rowlinson:1969}
\end{description}

\section{\label{FlucThermDerivs}Connecting fluctuations to thermodynamic derivatives}

If $A$ is a dynamical quantity which depends only on the
configurational degrees of freedom, then its average  in the canonical ensemble
(NVT) is given by (where, for convenience, we use a discrete-state notation,
with $A_i$ referring to the value of $A$ in the $i$th micro-state, etc.)

\begin{equation}
\langle A \rangle = \frac{\sum_i A_i \exp(-\beta U_i)}
{\sum_i \exp(-\beta U_i)} = \frac{\sum_i A_i \exp(-\beta U_i)}{Q}
\end{equation}

\nod where $\beta=1/k_BT$ and $Q$ is the configurational partition function. 
Then the 
inverse temperature derivative of $\langle A \rangle$ can be written
in terms of equilibrium fluctuations:

\begin{align}
\left(\frac{\partial\langle A \rangle}{\partial \beta} \right)_V = &  
-\frac{\sum_i A_i \exp(-\beta U_i) U_i}{Q} \\ & +
\frac{\sum_i A_i \exp(-\beta U_i) \sum_j \exp(-\beta U_j)
  U_j}{Q^2} \nonumber \\
=& -  \left( \langle A U \rangle -  \langle A\rangle
  \langle U \rangle \right) \\
=& - \langle \Delta A \Delta U \rangle.
\end{align}

\nod Now taking $A=W$ and $A=U$ successively we find that 

\begin{align}
\left(\frac{\partial\langle W \rangle}{\partial T} \right)_V
/ \left(\frac{\partial\langle  U \rangle}{\partial T}\right)_V
 & = \left(\frac{\partial\langle W \rangle}{\partial \beta}\right)_V  / 
\left(\frac{\partial\langle  U \rangle}{\partial \beta}\right)_V \nonumber \\
& = \frac{\langle \Delta W \Delta U \rangle}{\langle  (\Delta U)^2 \rangle}.
\label{derivativeLinearRegression}
\end{align}

\nod This last expression is precisely the formula for the slope obtained 
by linear-regression when plotting $W$ against $U$. 

Consider now volume 
derivatives. Because volume dependence comes in through the micro-state values,
$A_i$ and $U_i$, and the volume derivatives of these are not 
necessarily related in a simple way, the corresponding expression is not as 
simple: The derivative of $\angleb{W}$  with
respect to volume at fixed temperature is given by

\begin{align}
\left(\frac{\partial \angleb{W}}{\partial V}\right)_T & =
\frac{\partial}{\partial V}\Bigl(\angleb{p}V - Nk_BT\Bigr)_T \\ 
&= \angleb{p} + V\left(\frac{\partial \angleb{p}}{\partial V}\right)_T = \angleb{p} - K_T,
\end{align}

\nod where $K_T$ is the isothermal bulk modulus. The derivative of $U$
can be obtained by writing pressure as the derivative of Helmholtz
free energy $F$ ($K$ is kinetic energy):

\begin{align}
\angleb{p} & = -\left(\frac{\partial F}{\partial V}\right)_T =
-\left(\frac{\partial (\angleb{U}+\angleb{K}-TS)}{\partial V}\right)_T \\
&= -\left(\frac{\partial \angleb{U}}{\partial V}\right)_T + T
\left(\frac{\partial S}{\partial V}\right)_T.
\end{align}

\nod Then using the thermodynamic identity $\left(\frac{\partial S}
{\partial V}\right)_T = \left(\frac{\partial
 \angleb{p}}{\partial T}\right)_V\equiv 
\beta_V$, we obtain the ratio of volume derivatives of $\angleb{W}$ and
$\angleb{U}$

\begin{equation}
\left(\frac{\partial \angleb{W}}{\partial V}\right)_T / \left(\frac{\partial
 \angleb{U}}{\partial V}\right)_T =
-\frac{K_T-\angleb{p}}{T\beta_V - \angleb{p}}
\end{equation}

\nod which becomes $-K_T/(T\beta_V)$ when the pressure is small
compared to the bulk modulus. As discussed in Paper II, $\beta_V$ can be
expressed in terms of $\langle\Delta U\Delta W\rangle$ again, but the 
fluctuation expression for $K_T$ is more complicated. Thus we cannot simply
identify the lines of constant $T$, varying $V$, on a $\angleb{W},\angleb{U}$ 
plot, as we could with lines of fixed $V$, varying $T$, by examining the
fluctuations at fixed $V,T$.

\section{\label{Virial_SQW}Virial for square-well system}

\begin{figure}
\includegraphics[width=8.5cm]{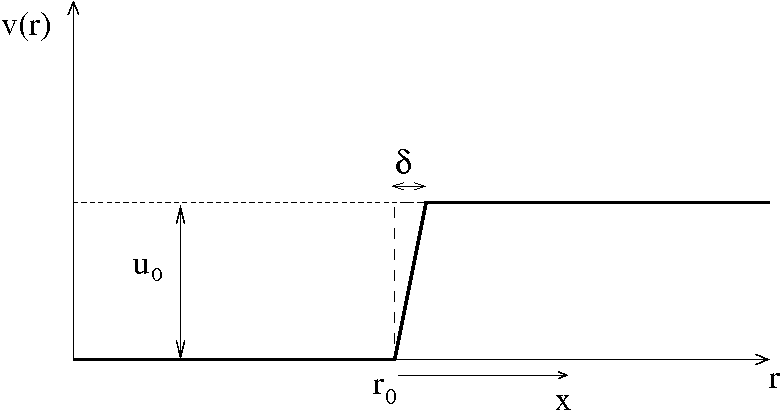}
\caption{\label{illustrateSQWvirial}Illustration of replacement of 
discontinuous step by a finite slope for the square well potential for the
purpose of calculating the virial. The limit $\delta\rightarrow 0$ is taken
at the end.}
\end{figure}

Here we derive the expression for the time-averaged virial, 
Eq.~(\ref{SQWvirial}), as a convenience for the reader. The idea is to replace 
the step $u_0$ in the potential with a
a finite slope $u_0/\delta$ over a range $\delta$, and take the 
limit $\delta \rightarrow 0$. We start by replacing a two-body interaction in
three dimensions with the equivalent one-dimensional, one-body problem using 
the radial separation $r$ and the reduced mass $m_r$. Let the potential step
be at $r=r_s$ and define $x=r-r_s$ (see Fig.~\ref{illustrateSQWvirial}). We 
consider an ``escape event'' over a 
positive step, so that an initial (relative) velocity $v_0$ becomes a final
velocity $v_1$ and $r$ goes from a value less than $r_0$ to a value greater than
$r_0+\delta$. The resulting formula
also applies for the other kinds of events.

The contribution to the time integral of the virial from this
event is given by

\begin{equation}
  \Delta  = \int_0^{t_\delta} \frac{(r_0+x)}{3} F dt =
  -\int_0^{t_\delta}  \frac{(r_0 + x)u_0}{3\delta} dt
\end{equation}

\nod where $F$ is the (constant) force in the region $0<x<\delta$ and $t_\delta$
is the time taken for the 'particle' (the radial separation) to cross this
region. The trajectory $x(t)$ is given by the standard formula for uniform
acceleration

\begin{equation}\label{x_of_t}
  x(t) = v_0 t - \half \frac{u_0}{\delta m_r} t^2,
\end{equation}

\nod which by setting $x(t_\delta)=\delta$ gives the following expression for
$t_\delta$:

\begin{equation}\label{t_delta}
  t_\delta = \delta \left( \frac{m_r v_0}{u_0} - \sqrt{\left( \frac{m_r v_0}
        {u_0}\right)^2 - \frac{2m_r}{u_0}} \right);
\end{equation}

\nod here we have taken the negative root, appropriate for a positive 
$u_0$ (we want the smallest positive $t_\delta$). Returning to 
$\Delta$, it can be split into two parts as follows

\begin{equation}
\Delta = -\int_0^{t_\delta} \frac{r_0 u_0}{3\delta} dt - \int_0^{t_\delta} \frac{x(t) u_0}{3\delta} dt
\end{equation}

Consider the second term: using the expression for $x(t)$ from 
Eq.~(\ref{x_of_t}), we see that the result of the integral will involve a term 
proportional to $t_\delta^2$ and one proportional to $t_\delta^3$. Using
Eq.~(\ref{t_delta}) to replace $t_\delta\propto \delta$, and noting the 
 $\delta$ in the denominator, the terms will have linear and quadratic
dependence on $\delta$, respectively. Thus they will vanish in the limit 
$\delta\rightarrow 0$. On the other hand, the first term gives

\begin{align}
\Delta = -\frac{r_0 u_0}{3\delta} t_\delta &= -\frac{r_0 u_0}{3}
\left( \frac{m_r v_0}{u_0} - \sqrt{\left( \frac{m_r v_0}
{u_0}\right)^2 - \frac{2m_r}{u_0}} \right)\nonumber \\
&= \frac{r_0 m_r}{3} \left( \sqrt{v_0^2 - 2 u_0/m_r} - v_0 \right).
\end{align}

This expression can be simplified by writing it in terms of the change of
velocity $\Delta v\equiv v_1 - v_0$. In the one-body problem conservation of 
momentum does not hold, and $v_1$ is given by energy conservation:

\begin{equation}
\half m_r v_0^2 = \half m_r v_1^2 + u_0
\end{equation}

\nod from which $\Delta v$ is obtained as

\begin{equation}
\Delta v \equiv v_1 - v_0 = \sqrt{{v_0}^2-2u_0/m_r} - v_0,
\end{equation}

\nod thus the expression for $\Delta $ becomes

\begin{equation}
\Delta = \frac{r_0 m_r}{3} \Delta v = \frac { m_r}{3} \bfa{r} \cdot \Delta 
\bfa{v}
\end{equation}

\nod where in the last expression a switch to three-dimensional notation was
made, recognizing that for central potentials $\Delta \bfa{v}$ will be parallel
to the displacement vector between the two particles. This expression,
derived for escape events, must also hold for capture events since these are
time-reverses of each other, and the virial is fundamentally a configurational
quantity, independent of dynamics (the above expression is
 time-reversal invariant because the change in the radial component of
velocity is the same either way, since although the ``initial'' and ``final'' 
velocities are swapped, they also have opposite sign). Bounce and collision
events may be treated by dividing the event into two parts at the turning point
(where the relative velocity is zero), noting that each may be treated exactly
as above, then adding the results back together. If we now consider 
all events that take place during an 
averaging time $t_{\textrm{av}}$), we get the time-averaged virial as

\begin{equation}
\bar{W}  = \frac{1}{3 t_{\textrm{av}}} \sum_{\textrm{events}\, e} m_{r,e}\bfa{r}_{e} 
\cdot \Delta \bfa{v}_{e}
\end{equation}


\begin{thebibliography}{32}
\expandafter\ifx\csname natexlab\endcsname\relax\def\natexlab#1{#1}\fi
\expandafter\ifx\csname bibnamefont\endcsname\relax
  \def\bibnamefont#1{#1}\fi
\expandafter\ifx\csname bibfnamefont\endcsname\relax
  \def\bibfnamefont#1{#1}\fi
\expandafter\ifx\csname citenamefont\endcsname\relax
  \def\citenamefont#1{#1}\fi
\expandafter\ifx\csname url\endcsname\relax
  \def\url#1{\texttt{#1}}\fi
\expandafter\ifx\csname urlprefix\endcsname\relax\def\urlprefix{URL }\fi
\providecommand{\bibinfo}[2]{#2}
\providecommand{\eprint}[2][]{\url{#2}}

\bibitem[{\citenamefont{Landau and Lifshitz}(1980)}]{Landau/Lifshitz:1980}
\bibinfo{author}{\bibfnamefont{L.~D.} \bibnamefont{Landau}} \bibnamefont{and}
  \bibinfo{author}{\bibfnamefont{E.~M.} \bibnamefont{Lifshitz}},
  \emph{\bibinfo{title}{{Statistical Physics, Part I}}}
  (\bibinfo{publisher}{Pergamon Press, London}, \bibinfo{year}{1980}).

\bibitem[{\citenamefont{Hansen and McDonald}(1986)}]{Hansen/McDonald:1986}
\bibinfo{author}{\bibfnamefont{J.~P.} \bibnamefont{Hansen}} \bibnamefont{and}
  \bibinfo{author}{\bibfnamefont{I.~R.} \bibnamefont{McDonald}},
  \emph{\bibinfo{title}{{Theory of Simple Liquids}}}
  (\bibinfo{publisher}{Academic Press, New York}, \bibinfo{year}{1986}),
  \bibinfo{edition}{2nd} ed.

\bibitem[{\citenamefont{Reichl}(1998)}]{Reichl:1998}
\bibinfo{author}{\bibfnamefont{L.~E.} \bibnamefont{Reichl}},
  \emph{\bibinfo{title}{{A Modern Course in Statistical Physics}}}
  (\bibinfo{publisher}{Wiley, New York}, \bibinfo{year}{1998}),
  \bibinfo{edition}{2nd} ed.

\bibitem[{\citenamefont{Allen and Tildesley}(1987)}]{Allen/Tildesley:1987}
\bibinfo{author}{\bibfnamefont{M.~P.} \bibnamefont{Allen}} \bibnamefont{and}
  \bibinfo{author}{\bibfnamefont{D.~J.} \bibnamefont{Tildesley}},
  \emph{\bibinfo{title}{Computer Simulation of Liquids}}
  (\bibinfo{publisher}{Oxford University Press}, \bibinfo{year}{1987}).

\bibitem[{\citenamefont{Bailey et~al.}(2008)\citenamefont{Bailey, Pedersen,
  Gnan, Schr{\o}der, and Dyre}}]{Bailey/others:2008c}
\bibinfo{author}{\bibfnamefont{N.~P.} \bibnamefont{Bailey}},
  \bibinfo{author}{\bibfnamefont{U.~R.} \bibnamefont{Pedersen}},
  \bibinfo{author}{\bibfnamefont{N.}~\bibnamefont{Gnan}},
  \bibinfo{author}{\bibfnamefont{T.~B.} \bibnamefont{Schr{\o}der}},
  \bibnamefont{and} \bibinfo{author}{\bibfnamefont{J.~C.} \bibnamefont{Dyre}}
  (\bibinfo{year}{2008}), \bibinfo{note}{to appear in J. Chem. Phys.}

\bibitem[{\citenamefont{Lennard-Jones}(1931)}]{Lennard-Jones:1931}
\bibinfo{author}{\bibfnamefont{J.~E.} \bibnamefont{Lennard-Jones}},
  \bibinfo{journal}{Proc. Phys. Soc. London} \textbf{\bibinfo{volume}{43}},
  \bibinfo{pages}{461} (\bibinfo{year}{1931}).

\bibitem[{\citenamefont{Weeks et~al.}(1971)\citenamefont{Weeks, Chandler, and
  Andersen}}]{Weeks/Chandler/Andersen:1971}
\bibinfo{author}{\bibfnamefont{J.~D.} \bibnamefont{Weeks}},
  \bibinfo{author}{\bibfnamefont{D.}~\bibnamefont{Chandler}}, \bibnamefont{and}
  \bibinfo{author}{\bibfnamefont{H.~C.} \bibnamefont{Andersen}},
  \bibinfo{journal}{J. Chem. Phys.} \textbf{\bibinfo{volume}{54}},
  \bibinfo{pages}{5237} (\bibinfo{year}{1971}).

\bibitem[{\citenamefont{Ben-Amotz and Stell}(2003)}]{Ben-Amotz/Stell:2003}
\bibinfo{author}{\bibfnamefont{D.}~\bibnamefont{Ben-Amotz}} \bibnamefont{and}
  \bibinfo{author}{\bibfnamefont{G.}~\bibnamefont{Stell}}, \bibinfo{journal}{J.
  Chem. Phys.} \textbf{\bibinfo{volume}{119}}, \bibinfo{pages}{10777}
  (\bibinfo{year}{2003}).

\bibitem[{\citenamefont{Pedersen
  et~al.}(2008{\natexlab{a}})\citenamefont{Pedersen, Bailey, Schr{\o}der, and
  Dyre}}]{Pedersen/others:2008}
\bibinfo{author}{\bibfnamefont{U.~R.} \bibnamefont{Pedersen}},
  \bibinfo{author}{\bibfnamefont{N.~P.} \bibnamefont{Bailey}},
  \bibinfo{author}{\bibfnamefont{T.~B.} \bibnamefont{Schr{\o}der}},
  \bibnamefont{and} \bibinfo{author}{\bibfnamefont{J.~C.} \bibnamefont{Dyre}},
  \bibinfo{journal}{Phys. Rev. Lett.} \textbf{\bibinfo{volume}{100}},
  \bibinfo{pages}{015701} (\bibinfo{year}{2008}{\natexlab{a}}).

\bibitem[{\citenamefont{Landau and Binder}(2005)}]{Landau/Binder:2005}
\bibinfo{author}{\bibfnamefont{D.~P.} \bibnamefont{Landau}} \bibnamefont{and}
  \bibinfo{author}{\bibfnamefont{K.}~\bibnamefont{Binder}},
  \emph{\bibinfo{title}{A Guide to Monte Carlo Simulations in Statistical
  Physics}} (\bibinfo{publisher}{Cambridge University Press},
  \bibinfo{year}{2005}), \bibinfo{edition}{2nd} ed.

\bibitem[{\citenamefont{Zaccarelli et~al.}(2002)\citenamefont{Zaccarelli,
  Foffi, Dawson, Buldyrev, Sciortino, and Tartaglia}}]{Zaccarelli/others:2002}
\bibinfo{author}{\bibfnamefont{E.}~\bibnamefont{Zaccarelli}},
  \bibinfo{author}{\bibfnamefont{G.}~\bibnamefont{Foffi}},
  \bibinfo{author}{\bibfnamefont{K.~A.} \bibnamefont{Dawson}},
  \bibinfo{author}{\bibfnamefont{S.~V.} \bibnamefont{Buldyrev}},
  \bibinfo{author}{\bibfnamefont{F.}~\bibnamefont{Sciortino}},
  \bibnamefont{and}
  \bibinfo{author}{\bibfnamefont{P.}~\bibnamefont{Tartaglia}},
  \bibinfo{journal}{Phys. Rev. E} \textbf{\bibinfo{volume}{66}},
  \bibinfo{pages}{041402} (\bibinfo{year}{2002}).

\bibitem[{\citenamefont{Berendsen et~al.}(1995)\citenamefont{Berendsen, van~der
  Spoel, and van Drunen}}]{Berendsen/vanderSpoel/vanDrunen:1995}
\bibinfo{author}{\bibfnamefont{H.~J.~C.} \bibnamefont{Berendsen}},
  \bibinfo{author}{\bibfnamefont{D.}~\bibnamefont{van~der Spoel}},
  \bibnamefont{and} \bibinfo{author}{\bibfnamefont{R.}~\bibnamefont{van
  Drunen}}, \bibinfo{journal}{Comp. Phys. Comm.} \textbf{\bibinfo{volume}{91}},
  \bibinfo{pages}{43} (\bibinfo{year}{1995}).

\bibitem[{\citenamefont{Lindahl et~al.}(2001)\citenamefont{Lindahl, Hess, and
  van~der Spoel}}]{Lindahl/Hess/vanderSpoel:2001}
\bibinfo{author}{\bibfnamefont{E.}~\bibnamefont{Lindahl}},
  \bibinfo{author}{\bibfnamefont{B.}~\bibnamefont{Hess}}, \bibnamefont{and}
  \bibinfo{author}{\bibfnamefont{D.}~\bibnamefont{van~der Spoel}},
  \bibinfo{journal}{J. Mol. Mod.} \textbf{\bibinfo{volume}{7}},
  \bibinfo{pages}{306} (\bibinfo{year}{2001}).

\bibitem[{Asap()}]{Asap}
Asap, \emph{\bibinfo{title}{{Asap} home page}}, \bibinfo{note}{{\tt
  https://wiki.fysik.dtu.dk/asap}},
  \urlprefix\url{https://wiki.fysik.dtu.dk/asap}.

\bibitem[{\citenamefont{Bailey et~al.}(2006)\citenamefont{Bailey, Cretegny,
  Sethna, Coffman, Dolgert, Myers, Schi{\o}tz, and
  Mortensen}}]{Bailey/others:2006}
\bibinfo{author}{\bibfnamefont{N.~P.} \bibnamefont{Bailey}},
  \bibinfo{author}{\bibfnamefont{T.}~\bibnamefont{Cretegny}},
  \bibinfo{author}{\bibfnamefont{J.~P.} \bibnamefont{Sethna}},
  \bibinfo{author}{\bibfnamefont{V.~R.} \bibnamefont{Coffman}},
  \bibinfo{author}{\bibfnamefont{A.~J.} \bibnamefont{Dolgert}},
  \bibinfo{author}{\bibfnamefont{C.~R.} \bibnamefont{Myers}},
  \bibinfo{author}{\bibfnamefont{J.}~\bibnamefont{Schi{\o}tz}},
  \bibnamefont{and} \bibinfo{author}{\bibfnamefont{J.~J.}
  \bibnamefont{Mortensen}} (\bibinfo{year}{2006}), \eprint{cond-mat/0601236}.

\bibitem[{\citenamefont{Jacobsen et~al.}(1987)\citenamefont{Jacobsen,
  N{\o}rskov, and Puska}}]{Jacobsen/Norskov/Puska:1987}
\bibinfo{author}{\bibfnamefont{K.~W.} \bibnamefont{Jacobsen}},
  \bibinfo{author}{\bibfnamefont{J.~K.} \bibnamefont{N{\o}rskov}},
  \bibnamefont{and} \bibinfo{author}{\bibfnamefont{M.~J.} \bibnamefont{Puska}},
  \bibinfo{journal}{Phys. Rev. B} \textbf{\bibinfo{volume}{35}},
  \bibinfo{pages}{7423} (\bibinfo{year}{1987}).

\bibitem[{\citenamefont{Jacobsen et~al.}(1996)\citenamefont{Jacobsen, Stoltze,
  and N{\o}rskov}}]{Jacobsen/Stoltze/Norskov:1996}
\bibinfo{author}{\bibfnamefont{K.~W.} \bibnamefont{Jacobsen}},
  \bibinfo{author}{\bibfnamefont{P.}~\bibnamefont{Stoltze}}, \bibnamefont{and}
  \bibinfo{author}{\bibfnamefont{J.~K.} \bibnamefont{N{\o}rskov}},
  \bibinfo{journal}{Surf. Sci.} \textbf{\bibinfo{volume}{366}},
  \bibinfo{pages}{394} (\bibinfo{year}{1996}).

\bibitem[{\citenamefont{Pedersen
  et~al.}(2008{\natexlab{b}})\citenamefont{Pedersen, Christensen, Schr{\o}der,
  and Dyre}}]{Pedersen/others:2008a}
\bibinfo{author}{\bibfnamefont{U.~R.} \bibnamefont{Pedersen}},
  \bibinfo{author}{\bibfnamefont{T.}~\bibnamefont{Christensen}},
  \bibinfo{author}{\bibfnamefont{T.~B.} \bibnamefont{Schr{\o}der}},
  \bibnamefont{and} \bibinfo{author}{\bibfnamefont{J.~C.} \bibnamefont{Dyre}},
  \bibinfo{journal}{Phys. Rev. E} \textbf{\bibinfo{volume}{77}},
  \bibinfo{pages}{011201} (\bibinfo{year}{2008}{\natexlab{b}}).

\bibitem[{\citenamefont{Dzugutov}(1992)}]{Dzugutov:1992}
\bibinfo{author}{\bibfnamefont{M.}~\bibnamefont{Dzugutov}},
  \bibinfo{journal}{Phys. Rev. A} \textbf{\bibinfo{volume}{46}},
  \bibinfo{pages}{R2984} (\bibinfo{year}{1992}).

\bibitem[{\citenamefont{Kob and Andersen}(1994)}]{Kob/Andersen:1994}
\bibinfo{author}{\bibfnamefont{W.}~\bibnamefont{Kob}} \bibnamefont{and}
  \bibinfo{author}{\bibfnamefont{H.~C.} \bibnamefont{Andersen}},
  \bibinfo{journal}{Phys. Rev. Lett.} \textbf{\bibinfo{volume}{73}},
  \bibinfo{pages}{1376} (\bibinfo{year}{1994}).

\bibitem[{\citenamefont{Scott et~al.}(1999)\citenamefont{Scott, Hunenberger,
  Tironi, Mark, Billeter, Fennen, Torda, Huber, Kruger, and van
  Gunsteren}}]{Scott/others:1999}
\bibinfo{author}{\bibfnamefont{W.~R.~P.} \bibnamefont{Scott}},
  \bibinfo{author}{\bibfnamefont{P.~H.} \bibnamefont{Hunenberger}},
  \bibinfo{author}{\bibfnamefont{I.~G.} \bibnamefont{Tironi}},
  \bibinfo{author}{\bibfnamefont{A.~E.} \bibnamefont{Mark}},
  \bibinfo{author}{\bibfnamefont{S.~R.} \bibnamefont{Billeter}},
  \bibinfo{author}{\bibfnamefont{J.}~\bibnamefont{Fennen}},
  \bibinfo{author}{\bibfnamefont{A.~E.} \bibnamefont{Torda}},
  \bibinfo{author}{\bibfnamefont{T.}~\bibnamefont{Huber}},
  \bibinfo{author}{\bibfnamefont{P.}~\bibnamefont{Kruger}}, \bibnamefont{and}
  \bibinfo{author}{\bibfnamefont{W.}~\bibnamefont{van Gunsteren}},
  \bibinfo{journal}{J. Phys. Chem. A} \textbf{\bibinfo{volume}{103}},
  \bibinfo{pages}{3596} (\bibinfo{year}{1999}).

\bibitem[{\citenamefont{Bailey et~al.}(2004)\citenamefont{Bailey, Schi{\o}tz,
  and Jacobsen}}]{Bailey/Schiotz/Jacobsen:2004a}
\bibinfo{author}{\bibfnamefont{N.~P.} \bibnamefont{Bailey}},
  \bibinfo{author}{\bibfnamefont{J.}~\bibnamefont{Schi{\o}tz}},
  \bibnamefont{and} \bibinfo{author}{\bibfnamefont{K.~W.}
  \bibnamefont{Jacobsen}}, \bibinfo{journal}{Phys. Rev. B}
  \textbf{\bibinfo{volume}{69}}, \bibinfo{pages}{144205}
  (\bibinfo{year}{2004}).

\bibitem[{\citenamefont{Lewis and Wahnstr{\"o}m}(1994)}]{Lewis/Wahnstrom:1994}
\bibinfo{author}{\bibfnamefont{L.~J.} \bibnamefont{Lewis}} \bibnamefont{and}
  \bibinfo{author}{\bibfnamefont{G.}~\bibnamefont{Wahnstr{\"o}m}},
  \bibinfo{journal}{Phys. Rev. E} \textbf{\bibinfo{volume}{50}},
  \bibinfo{pages}{3865} (\bibinfo{year}{1994}).

\bibitem[{\citenamefont{Berendsen et~al.}(1987)\citenamefont{Berendsen,
  Grigera, and Straatsma}}]{Berendsen/Grigera/Straatsma:1987}
\bibinfo{author}{\bibfnamefont{H.~J.~C.} \bibnamefont{Berendsen}},
  \bibinfo{author}{\bibfnamefont{J.~R.} \bibnamefont{Grigera}},
  \bibnamefont{and} \bibinfo{author}{\bibfnamefont{T.~P.}
  \bibnamefont{Straatsma}}, \bibinfo{journal}{J. Phys. Chem.}
  \textbf{\bibinfo{volume}{91}}, \bibinfo{pages}{6269} (\bibinfo{year}{1987}).

\bibitem[{\citenamefont{Zaccarelli et~al.}(2004)\citenamefont{Zaccarelli,
  Sciortino, and Tartaglia}}]{Zaccarelli/Sciortino/Tartaglia:2004}
\bibinfo{author}{\bibfnamefont{E.}~\bibnamefont{Zaccarelli}},
  \bibinfo{author}{\bibfnamefont{F.}~\bibnamefont{Sciortino}},
  \bibnamefont{and}
  \bibinfo{author}{\bibfnamefont{P.}~\bibnamefont{Tartaglia}},
  \bibinfo{journal}{J. Phys.: Condens. Matt.} \textbf{\bibinfo{volume}{16}},
  \bibinfo{pages}{4849} (\bibinfo{year}{2004}).

\bibitem[{\citenamefont{Mahoney and Jorgensen}(2000)}]{Mahoney/Jorgensen:2000}
\bibinfo{author}{\bibfnamefont{M.~W.} \bibnamefont{Mahoney}} \bibnamefont{and}
  \bibinfo{author}{\bibfnamefont{W.~L.} \bibnamefont{Jorgensen}},
  \bibinfo{journal}{J. Chem. Phys.} \textbf{\bibinfo{volume}{112}},
  \bibinfo{pages}{8910} (\bibinfo{year}{2000}).

\bibitem[{\citenamefont{Sciortino}(2002)}]{Sciortino:2002}
\bibinfo{author}{\bibfnamefont{F.}~\bibnamefont{Sciortino}},
  \bibinfo{journal}{Nature Mater.} \textbf{\bibinfo{volume}{1}},
  \bibinfo{pages}{145} (\bibinfo{year}{2002}).

\bibitem[{\citenamefont{Daw and Baskes}(1984)}]{Daw/Baskes:1984}
\bibinfo{author}{\bibfnamefont{M.~R.} \bibnamefont{Daw}} \bibnamefont{and}
  \bibinfo{author}{\bibfnamefont{M.~I.} \bibnamefont{Baskes}},
  \bibinfo{journal}{Phys. Rev. B} \textbf{\bibinfo{volume}{29}},
  \bibinfo{pages}{6443} (\bibinfo{year}{1984}).

\bibitem[{\citenamefont{Hansen and McDonald}(1969)}]{Rowlinson:1969}
\bibinfo{author}{\bibfnamefont{J.~P.} \bibnamefont{Hansen}} \bibnamefont{and}
  \bibinfo{author}{\bibfnamefont{I.~R.} \bibnamefont{McDonald}},
  \emph{\bibinfo{title}{{Liquid and Liquid Mixtures}}}
  (\bibinfo{publisher}{Butterworths, London}, \bibinfo{year}{1969}).

\bibitem[{\citenamefont{van Gunsteren et~al.}(1996)\citenamefont{van Gunsteren,
  Billeter, Eising, H{\"u}nenberger, Kr\"uger, Mark, Scott, and
  Tironi}}]{vanGunsteren/others:1996}
\bibinfo{author}{\bibfnamefont{W.~F.} \bibnamefont{van Gunsteren}},
  \bibinfo{author}{\bibfnamefont{S.~R.} \bibnamefont{Billeter}},
  \bibinfo{author}{\bibfnamefont{A.~A.} \bibnamefont{Eising}},
  \bibinfo{author}{\bibfnamefont{P.~H.} \bibnamefont{H{\"u}nenberger}},
  \bibinfo{author}{\bibfnamefont{P.}~\bibnamefont{Kr\"uger}},
  \bibinfo{author}{\bibfnamefont{A.~E.} \bibnamefont{Mark}},
  \bibinfo{author}{\bibfnamefont{W.~R.~P.} \bibnamefont{Scott}},
  \bibnamefont{and} \bibinfo{author}{\bibfnamefont{I.~G.}
  \bibnamefont{Tironi}}, \emph{\bibinfo{title}{{Biomolecular Simulation: The
  GROMOS96 manual and user guide}}} (\bibinfo{publisher}{Hochschulverlag AG an
  der ETH Z\"urich, Z\"urich, Switzerland}, \bibinfo{year}{1996}).

\bibitem[{\citenamefont{Jorgensen et~al.}(1984)\citenamefont{Jorgensen, Madura,
  and Swenson}}]{Jorgensen/Madura/Swenson:1984}
\bibinfo{author}{\bibfnamefont{W.~L.} \bibnamefont{Jorgensen}},
  \bibinfo{author}{\bibfnamefont{J.~D.} \bibnamefont{Madura}},
  \bibnamefont{and} \bibinfo{author}{\bibfnamefont{C.~J.}
  \bibnamefont{Swenson}}, \bibinfo{journal}{J. Am. Chem. Soc.}
  \textbf{\bibinfo{volume}{106}}, \bibinfo{pages}{6638} (\bibinfo{year}{1984}).

\bibitem[{\citenamefont{Esbensen et~al.}(2002)\citenamefont{Esbensen, Guyot,
  Westad, and Houm{\o}ller}}]{Esbensen/others:2002}
\bibinfo{author}{\bibfnamefont{K.~H.} \bibnamefont{Esbensen}},
  \bibinfo{author}{\bibfnamefont{D.}~\bibnamefont{Guyot}},
  \bibinfo{author}{\bibfnamefont{F.}~\bibnamefont{Westad}}, \bibnamefont{and}
  \bibinfo{author}{\bibfnamefont{L.~P.} \bibnamefont{Houm{\o}ller}},
  \emph{\bibinfo{title}{{Multivariate Data Analysis - In practice}}}
  (\bibinfo{publisher}{Camo, Oslo}, \bibinfo{year}{2002}),
  \bibinfo{edition}{5th} ed.

\end{thebibliography}
\end{document}